\documentclass[a4paper,10pt]{article}

\usepackage{amsmath}
\usepackage{amsthm}
\usepackage{amsfonts}
\usepackage{hyperref}
\usepackage{color}
\usepackage{multirow}
\usepackage[dvips]{graphicx}
\usepackage{subfigure}
\usepackage{amssymb}
\usepackage{epsfig}

\newcommand{\comment}[1]{}
\numberwithin{equation}{section}

\begin{document}

\bibliographystyle{plain}

\title{\textbf{
Wetting Boundary Conditions in Phase-Field-Based Simulation of Binary Fluids: 
Some Comparative Studies and New Development
}}

\author{$\textbf{Jun-Jie Huang}^{1, 2,  3\footnote{Corresponding author. E-mail: jjhuang1980@gmail.com; jjhuang@cqu.edu.cn.}}, 
\textbf{Haibo Huang}^{4},
\textbf{Xinzhu Wang}^{1, 2}$\\
\\
$ ^1$  Department of Engineering Mechanics, \\
Chongqing University, Chongqing 400044, China \\
$ ^2$ Chongqing Key Laboratory of Heterogeneous Material Mechanics \\
(Chongqing University), Chongqing 400044, China\\
$ ^3$ State Key Laboratory of Mechanical Transmission, \\
Chongqing University, Chongqing 400044, China \\
$ ^4$ Department of Modern Mechanics, \\
University of Science and Technology of China, Hefei, Anhui 230026, China}

\maketitle

\textbf{Abstract}

We studied several wetting boundary conditions (WBCs) in the simulation of binary fluids
based on phase-field theory.
Five WBCs, three belonging to the surface energy (SE) formulation using the linear, cubic
and sine functions (denoted as LinSE, CubSE and SinSE), 
the fourth using a geometric formulation (Geom), 
and the fifth using a characteristic interpolation (CI),
were compared with each other through the study of several problems:
(1) the static contact angle of a drop; 
(2) a Poiseuille flow-driven liquid column;
(3) a wettability gradient (WG)-driven liquid column;
(4) drop dewetting.
It was found that while all WBCs can predict the static contact angle fairly accurately, 
they may affect the simulation outcomes of dynamic problems differently, depending on the driving mechanism. 
For the flow-driven problem, to use different WBCs had almost no effect
on the flow characteristics over a large scale.
But for other capillarity-driven problems, the WBC had some noticeable effects.
For the WG-driven liquid column, Geom gave the most consistent prediction 
between the drop velocity and dynamic contact angles,
and LinSE delivered the poorest prediction in this aspect.
Except for Geom, 
the dynamic contact angle differed from the prescribed (static) one when other WBCs were used.
For drop dewetting, Geom led to the most violent drop motion 
whereas CubSE caused the weakest motion; 
the initial contact line velocity was also found to be dependent on the WBC.
For several problems, CubSE and SinSE gave almost the same results,
and those by Geom and CI were close as well,
possibly due to similar consideration in their design.
Besides various comparisons, a new implementation that may be used for all WBCs
was proposed to mimic the wall energy relaxation and control the degree of slip. 
This new procedure made it possible to allow the simulations  
to match experimental measurements well.

\textit{Wetting Boundary Condition}, 
\textit{Surface Energy},
\textit{Contact Angle}, 
\textit{Drop Simulation}.

\newpage

\section{Introduction}\label{sec:intro}

Two-phase flows near solid walls are encountered in our daily life,
in many industries, and also in some new technologies like
\textit{lab-on-a-chip}.
Computer simulations of such flows have gradually become mature in the
past few decades.
The popular simulation methods include the front-tracking (FT)~\cite{FrontTracking1992},
volume-of-fluid (VOF)~\cite{VOFReview99}, 
level-set (LS)~\cite{jcp96:ls}, and phase-field (PF) / diffuse-interface methods~\cite{DIMReview98, jacqmin99jcp}.
For isothermal near-wall two-phase flows, there are two additional
key issues in their computer simulations as compared with single-phase flows:
(1) to handle properly the  interface motion and the coupling between
the interfacial tension effects and the flow;
(2) to handle the wetting of the fluids on the solid wall and the
motion of contact lines.
Among various simulation methods, the PF-based one has solid
physical basis in dealing with the interface and wetting since it is
closely connected with the theory for fluids near critical
points~\cite{DIMReview98, cahnhilliard58}.
It is also a popular method in material science~\cite{pfssdewetting} and has been applied
in other challenging fields like complex
fluids~\cite{jcp06pf-fem-adaptive} 
and tumor growth modeling~\cite{jmb09pftumorgrowth}.
Under the general phase-field or diffuse-interface framework, 
there exist two approaches according to the way to solve
the governing equations: one directly solves the Navier-Stokes
equations (NSEs) and the interface evolution equation (usually the
Cahn-Hilliard equation (CHE)),
and the other is the free energy-based lattice Boltzmann method (LBM)
~\cite{Chen1998, succi:lbebook, swift95, swift96}, which solves the
evolution equations of some particle distribution functions.
The two approaches differ from each other in a few aspects, but the key components
related to interface and wetting modeling are similar or even exactly
the same. In this work the way to solve the governing equations is not a major
concern, and we will not discuss such differences in the following.
Our main focus is the wetting condition on a wall.

For small scale near-wall two-phase flows, which usually have low
Reynolds and capillary numbers, it is especially important to
capture the dynamics of the interfaces and the flows near the wall accurately.
Because of this requirement, the wetting condition on a wall has become a
key issue in the PF-based simulation of such flows
(it is also important in other methods like the VOF and LS methods, but we
concentrate on the PF-based here).
There are several kinds of implementations of the WBC based on
different considerations.
Many early studies employed the surface energy (SE) formulation with a
linear form SE density (denoted as LinSE)
~\cite{tipm02, briant02:mpwbc, FELBM-CL04a, FELBM-CL04b},
although the cubic form SE density (denoted as CubSE) 
was considered and used even earlier by
Jacqmin~\cite{jacqmin00jfm}.
Because CubSE avoids the appearance of the \textit{wall
  layer} (a layer enriched with one of the fluids while depleted in
the other)~\cite{jacqmin00jfm, ijmpc09wall-fe-bc}, it has been
employed by many others~\cite{ijmf06-cap-dri-flows, DIMSpreading07,
jfm10-sil-che-cl, cpc11wbc-lbm, jcp10lbm-drop-impact, jast-hybrid-lbm-fvm}.
With additional phenomenological parameter introduced,
even more complicated (yet more sophisticated) WBCs using CubSE have been
proposed and employed by Carlson et al.~\cite{pof09-dyn-wetting}
and by Yue and Feng~\cite{pof11-wall-energy-relax}
(we note that this more general WBC was already discussed much earlier by
Jacqmin although he did not actually use it~\cite{jacqmin00jfm}).
Qian et al. used a somewhat different form of SE which employed a sine function~\cite{pre03-gnbc}
(denoted as SinSE).
Both CubSE and SinSE can ensure that the normal gradient of the order parameter
(nearly) vanishes in the bulk region of each fluid.
Another different approach is the geometric formulation proposed by
Ding and Spelt~\cite{pre07-geom-wbc} (denoted as Geom). 
The special feature of this
formulation is that the local
microscopic contact angle is always enforced. 
It has been employed for contact angle hysteresis modeling 
by Ding and Spelt (in PF simulation)~\cite{jfm08-drop-shear}, 
and by Wang et al. (in LBM simulation)~\cite{pre13cah-lbm}.
Yet another recent development of WBC was proposed by Lee and Kim~\cite{cf11-accur-cabc}
based on a characteristic interpolation (denoted as CI),
which was claimed to possess certain numerical advantages.
It actually resembles the geometric formulation 
because in essence it also tries to enforce the microscopic contact angle 
on the wall (but via a somewhat different means).

In the literature, there have been some studies on different WBCs.
Ding and Spelt~\cite{pre07-geom-wbc} compared the SE and
geometric formulations for an axisymmetric drop spreading on a
homogeneous surface and a three-dimensional (3-D) drop subject to a
shear flow. However, they only considered one particular form of SE
formulation, and the problems they studied were limited as well.
Liu and Lee~\cite{ijmpc09wall-fe-bc} compared three forms 
(linear, quadratic and cubic) of SE. But they focused on non-ideal gas
(single-component fluid) rather than bindary fluids, and they studied \textit{static cases only}. 
Wikland et al.~\cite{cpc11wbc-lbm} compared both LinSE and CubSE
for a static drop on a homogenous wall and also for the capillary intrusion problem. 
But they did not consider the geometric formulation 
and their problem settings were limited as well
(only 2-D cases were studied).
In view of many versions of WBC and the variety of drop problems,
it is necessary and worthy to examine various WBCs under more situations
so as to gain useful insights on their similarities and differences for future simulations of similar problems.
Besides, the inclusion of wall energy relaxation in SE formulation
has been very effective to improve the simulation results noticeablely
~\cite{pof09-dyn-wetting, pof11-wall-energy-relax}.
One may become curious whether it is possible to bring this (or a similar) concept 
into other kinds of WBCs like Geom or CI to achieve similar improvement.
Thus, the purpose of the present work is to further broaden our
understanding of the effects of WBC on near-wall drop simulations
by comparing different WBCs, 
including the SE formulations (LinSE, CubSE and SinSE), 
the geometric formulation and the CI-based WBC, 
for a few typical drop problems in 2-D and/or axisymmetric geometry.
And the feasibility of including wall energy relaxation in a simple way 
for WBCs other than CubSE will also be demonstrated.
The numerical method employed from the present study is 
a hybrid lattice-Boltzmann finite-difference method
recently developed~\cite{hybrid-mrt-lb-fd-axisym} that can be
used for axisymmetric two-phase flow problems.

The paper is organized as follows.
In Section~\ref{sec:method},
the phase-field model for binary fluids and 
the five WBCs on a wall to be studied are described. 
The numerical method, originally presented in Ref.~\cite{hybrid-mrt-lb-fd-axisym}, 
is also very briefly described in this section.
In Section~\ref{sec:res-dis},
several drop problems, 
including the static contact angle of a drop, 
a flow-driven liquid column,
a liquid column driven by a stepwise WG,
and drop dewetting on a lyophobic wall, 
are investigated with different WBCs or their variants, 
and the respective results are compared and discussed.
Section~\ref{sec:conclusion} summarizes the findings and concludes this paper.

\section{Phase-Field Model, Wetting Boundary Condition 
and Numerical Method}\label{sec:method} 

\subsection{Phase-Field Model}\label{sec:pfm-if}

In the phase-field model, different fluids are distinguished by an
order parameter $\phi$,
based on which a free energy functional $\mathcal{F}$ is defined as,
\begin{equation}
  \label{eq:fe-functional-def}
\mathcal{F} (\phi, \boldsymbol{\nabla} \phi)
 = \int_{V} \bigg( \Psi (\phi) 
+ \frac{1}{2} \kappa \vert \boldsymbol{\nabla} \phi \vert ^2 
 \bigg) dV 
+ \int_{S} \varphi (\phi_{S} ) dS ,
\end{equation}
where $\Psi (\phi)$ is the \textit{bulk free energy} density 
and takes the following double-well form,
\begin{equation}
  \label{eq:bulk-fe}
  \Psi (\phi) = a (\phi^{2} - 1)^{2} ,
\end{equation}
with $a$ being a constant.
This form indicates that $\phi$ varies between $-1$ in one of the fluids
and $1$ in the other fluid. 
(Note that in some works the \textit{concentration} $C$,
instead of the \textit{order parameter} $\phi$,
is used and $C$ varies between $0$ and $1$;
but the two formulations using $\phi$ and $C$ can be converted into
each other through a linear transformation, thus there are no essential
differences between them.)
In Eq. (\ref{eq:fe-functional-def})
the second term is the \textit{interfacial energy} density
with $\kappa$ being another constant,
and the last term in the surface integral, $\varphi (\phi_{S} )$, is the 
\textit{surface energy} (SE) density 
with $\phi_{S}$ being the order parameter on the surface.
The specific form of $\varphi (\phi_{S} )$ will be discussed later.

By taking the variation of the free energy functional $\mathcal{F}$
with respect to the order parameter $\phi$,
one obtains the chemical potential $\mu$ as,
\begin{equation}
  \label{eq:chem-potential}
\mu = \frac{\delta \mathcal{F}}{\delta \phi} 
= \frac{d \Psi (\phi)}{d \phi} -  \kappa \nabla ^2 \phi
= 4 a \phi (\phi ^2 -1) -  \kappa \nabla ^2 \phi.
\end{equation}
The coefficients $a$ in the bulk free energy and 
$\kappa$ in the interfacial energy
are related to the interfacial tension
$\sigma$ and interface width $W$ as
~\cite{ijnmf09-pflbm-mobility},
\begin{equation}
  \label{eq:a-kappa-sigma-W}
a = \frac{3 \sigma}{4 W}, \quad
\kappa = \frac{3 \sigma W}{8}.
\end{equation}
Equivalently, the interfacial tension
$\sigma$ and interface width $W$ can be expressed in terms of $a$ and
$\kappa$ as,
\begin{equation}
  \label{eq:sigma-W-a-kappa}
\sigma  = \frac{4}{3} \sqrt{ 2 \kappa a}, \quad 
W = \sqrt{\frac{2 \kappa}{a}}.
\end{equation}

By assuming that the diffusion of the order parameter is driven by the
gradient of the chemical potential,
one has the following evolution equation for $\phi$~\cite{jacqmin99jcp},
\begin{equation}
  \label{eq:che}
\frac{\partial \phi}{\partial t}
+ (\boldsymbol{u} \cdot \boldsymbol{\nabla}) \phi
= \boldsymbol{\nabla} \cdot (M  \boldsymbol{\nabla} \mu)   ,
\end{equation}
where $M$ is the mobility (i.e., the diffusion coefficient, which is
taken as a constant here). Eq. (\ref{eq:che}) is known as the (convective)
Cahn-Hilliard equation (CHE).

Eqs. (\ref{eq:chem-potential}) and (\ref{eq:che}) must be supplemented
with certain boundary conditions. Here we only discuss the conditions near a wall 
which are closely related to the wetting and contact line.
As seen in Eq. (\ref{eq:che}), the fluid velocity also appears.
In this work, we assume the no-slip condition for the fluid velocity
$\boldsymbol{u}$ on a wall
and focus on the conditions for the PF variables.
It should be noted that in PF simulations the interface slip on a wall
is allowed because of the diffusion in Eq. (\ref{eq:che}).

\subsection{Wetting Boundary Condition}\label{ssec:wbc}

Near a solid wall, suitable boundary conditions are required for 
both the order parameter $\phi$ and the chemical potential $\mu$.
Different WBCs differ only in the condition for $\phi$,
while they share the same condition for $\mu$. 
The boundary condition on a surface for the chemical potential $\mu$
is the no-flux condition given by,
\begin{equation}
  \label{eq:bc-chem-potential}
  \boldsymbol{n}_{w} \cdot \boldsymbol{\nabla} \mu \vert _{S}
  = \frac{\partial \mu}{\partial n_{w}} \bigg\vert _{S} 
 = 0  ,
\end{equation}
where $\boldsymbol{n}_{w}$ denotes the unit normal vector on the surface
pointing into the fluid.
The different part (i.e., the condition for $\phi$)
is described for the five WBCs considered in this work as follows.

\subsubsection{WBC with the linear SE}\label{sssec:wbc-LinSE}

When the linear SE (LinSE) is used, 
one has the following form of
$\varphi (\phi_{S})$ in Eq. (\ref{eq:fe-functional-def}),
\begin{equation}
  \label{eq:LinSE}
\varphi (\phi_{S}) = - \omega \phi_{S} ,
\end{equation}
where $\omega$ is a parameter related to the wetting property of the surface.
Young's equation determines the (static) contact angle $\theta_{w}$ on the wall
(measured in fluid 1 with $\phi = 1$) as,
\begin{equation}
  \label{eq:CA-LinSE}
\cos \theta_{w} = \frac{1}{2} [  (\sqrt{1 + \tilde{\omega}}) ^{3} - (\sqrt{1 - \tilde{\omega}}) ^{3}] ,
\end{equation}
where the dimensionless parameter $\tilde{\omega}$ is defined as,
\begin{equation}
\tilde{\omega} = \frac{\omega}{\sqrt{2 \kappa a}} .
\end{equation}
The boundary condition for the order parameter $\phi$ (the natural
boundary condition) reads~\cite{FELBM-CL04b},
\begin{equation}
  \label{eq:bc-op-LinSE}
  \kappa \boldsymbol{n}_{w} \cdot \boldsymbol{\nabla} \phi \vert _{S}
  = \kappa \frac{\partial \phi}{\partial n_{w}} \bigg\vert _{S} 
  = \frac{d \varphi (\phi)}{d \phi} = - \omega = -  \sqrt{2 \kappa a} \tilde{\omega} .
\end{equation}
By using Eq. (\ref{eq:sigma-W-a-kappa}), one finds,
\begin{equation}
  \label{eq:bc-op-LinSE-simplified}
 \frac{\partial \phi}{\partial n_{w}} \bigg\vert _{S} 
  = -  \sqrt{\frac{2 a}{\kappa}} \tilde{\omega} = - \frac{1}{W/2} \tilde{\omega} .
\end{equation}

\subsubsection{WBC with the cubic SE}\label{sssec:wbc-cubse}

When the cubic SE (CubSE) is used, 
one has $\varphi (\phi_{S})$ in 
the following form~\cite{jfm10-sil-che-cl},
\begin{equation}
  \label{eq:CubSE}
\varphi (\phi_{S} ) = - \sigma \cos \theta_{w}
\frac{\phi_{S} (3 - \phi_{S}^{2})}{4} 
+ \frac{1}{2} (\sigma_{w1} + \sigma_{w2})  ,
\end{equation}
where $\varphi (\pm 1)$ gives the fluid-solid interfacial tensions
$\sigma_{w1}$ and $\sigma_{w2}$ between the wall
and fluid 1 (with $\phi = 1$) and fluid 2 (with $\phi = -1$), respectively. 
Similarly, Young's equation determines $\theta_{w}$ as,
\begin{equation}
  \cos \theta_{w} = \frac{\sigma_{w2} - \sigma_{w1}}{\sigma}  .
\end{equation}
Now the boundary condition for the order parameter $\phi$ 
reads~\cite{jfm10-sil-che-cl} (note that due to different definitions
of the unit normal vector $\boldsymbol{n}_{w}$, 
there is a change in the sign here as compared
with Ref.~\cite{jfm10-sil-che-cl}),
\begin{equation}
  \label{eq:bc-op-CubSE}
  \kappa \boldsymbol{n}_{w} \cdot \boldsymbol{\nabla} \phi \vert _{S}
  = \kappa \frac{\partial \phi}{\partial n_{w}} \bigg\vert _{S} 
 = \frac{d \varphi (\phi)}{d \phi}  
= - \frac{3 \sigma}{4} \cos \theta_{w} (1 - \phi_{S}^{2}) .
\end{equation}
By using Eq. (\ref{eq:a-kappa-sigma-W}), one finds,
\begin{equation}
  \label{eq:bc-op-CubSE-simplified}
 \frac{\partial \phi}{\partial n_{w}} \bigg\vert _{S} 
= - \frac{1}{W/2} \cos \theta_{w} (1 - \phi_{S}^{2}) .
\end{equation}
When Eq. (\ref{eq:bc-op-CubSE-simplified}) is compared with
Eq. (\ref{eq:bc-op-LinSE-simplified}), it is found that the constant
for LinSE $\tilde{\omega}$ is replaced by 
$\cos \theta_{w} (1 - \phi_{S}^{2})$ 
for CubSE (which is a function of $\phi_{S}$).
Therefore, the WBC using CubSE is somewhat more complicated than that
with LinSE and its implementation requires the order parameter on the
surface, $\phi_{S}$, which has to be found by some means first.

\subsubsection{WBC using a sine function SE}\label{sssec:wbc-sinse}

As mentioned earlier, in the literature
there is another form of SE proposed in~\cite{pre03-gnbc}
that uses a sine function (SinSE),
\begin{equation}
  \label{eq:SinSE}
\varphi (\phi_{S} ) =- \frac{\sigma}{2}  \cos \theta_{w}
\sin \bigg(  \frac{\pi}{2} \phi_{S} \bigg) .
\end{equation}
Now the boundary condition for the order parameter $\phi$ reads,
\begin{equation}
  \label{eq:bc-op-SinSE}
  \kappa \boldsymbol{n}_{w} \cdot \boldsymbol{\nabla} \phi \vert _{S}
  = \kappa \frac{\partial \phi}{\partial n_{w}} \bigg\vert _{S} 
 = \frac{d \varphi (\phi)}{d \phi}  
= - \frac{\sigma}{2} \cos \theta_{w} 
\bigg[ \frac{\pi}{2} \cos \bigg(  \frac{\pi}{2} \phi_{S} \bigg) \bigg].
\end{equation}
By using Eq. (\ref{eq:a-kappa-sigma-W}), one finds,
\begin{equation}
  \label{eq:bc-op-SinSE-simplified}
 \frac{\partial \phi}{\partial n_{w}} \bigg\vert _{S} 
= - \frac{1}{W/2} \cos \theta_{w} 
\bigg[ \frac{\pi}{3} \cos \bigg(  \frac{\pi}{2} \phi_{S} \bigg) \bigg] .
\end{equation}
When Eq. (\ref{eq:bc-op-SinSE-simplified}) is compared with
Eq. (\ref{eq:bc-op-LinSE-simplified}), it is found that the constant
for LinSE $\tilde{\omega}$ is replaced by 
$\cos \theta_{w} [ \frac{\pi}{3} \cos (  \frac{\pi}{2} \phi_{S} ) ]$ 
for SinSE (which is also a function of $\phi_{S}$).
Therefore, like above for CubSE, 
the WBC using SinSE is also more complicated than that
with LinSE and its implementation also requires 
to find the order parameter on the surface ($\phi_{S}$) by certain means.

It is seen from Eqs. (\ref{eq:bc-op-CubSE-simplified})
and (\ref{eq:bc-op-SinSE-simplified}) that the conditions for CubSE and SinSE
are actually quite similar and the only difference is in the function
on the right hand side (RHS).
Denote $f_{s1} (\phi) = (1 - \phi_{S}^{2})$ and 
$f_{s2} (\phi) =  \frac{\pi}{3} \cos (  \frac{\pi}{2} \phi_{S} )$.
It is not difficult to find out 
that both $f_{s1} (\phi)$ and $f_{s2} (\phi)$ satisfy the following conditions,
\begin{equation}
  \label{eq:CubSE-SinSE-common-cond}
f_{s} (\phi) \geq 0  \ (\textrm{for} \ -1 \leq \phi \leq 1 ), 
\quad f_{s} (1) = f_{s} (-1) = 0, \quad \int_{-1}^{1} f_{s} (\phi) d \phi = \frac{4}{3}  .
\end{equation}
When one plots the two functions for $-1 \leq \phi \leq 1$,
it is easy to see that they appear to be rather \textit{close} to each other.
Thus, it is expected that they would give results that are close as well
(which will be examined later).

\subsubsection{WBC in geometric formulation and the CI-based WBC}\label{sssec:wbc-geom}

The WBC in geometric formulation (Geom)
differs significantly from the
SE formulation presented above.
It abandons the surface energy integral and starts from some
geometric considerations. Specifically, it assumes that
the contours of the order parameter in the diffuse interface are
parallel to each other, including in the region near
the surface. Then, the normal vector to the interface, denoted by
$\boldsymbol{n}_{s}$, can be written in terms of the gradient of $\phi$ as~\cite{pre07-geom-wbc},
\begin{equation}
  \boldsymbol{n}_{s} = \frac{\boldsymbol{\nabla} \phi}{\vert \boldsymbol{\nabla} \phi\vert} .
\end{equation}
By taking note that $\phi$'s gradient may be decomposed as,
\begin{equation}
  \boldsymbol{\nabla} \phi = (\boldsymbol{n}_{w} \cdot \boldsymbol{\nabla}
  \phi) \boldsymbol{n}_{w} 
+ (\boldsymbol{t}_{w} \cdot \boldsymbol{\nabla} \phi) \boldsymbol{t}_{w}   ,
\end{equation}
where $\boldsymbol{t}_{w}$ is the unit tangential vector along the surface, one
finds that at the contact line the contact angle can be expressed by,
\begin{equation}
  \label{eq:geom-ca}
  \tan \bigg( \frac{\pi}{2}-\theta_{w} \bigg) 
= \frac{- \boldsymbol{n}_{w} \cdot \boldsymbol{\nabla} \phi}{\vert \boldsymbol{\nabla} \phi 
- (\boldsymbol{n}_{w} \cdot \boldsymbol{\nabla} \phi) \boldsymbol{n}_{w}
\vert} 
= \frac{- \boldsymbol{n}_{w} \cdot \boldsymbol{\nabla}
  \phi}{\vert (\boldsymbol{t}_{w} \cdot \boldsymbol{\nabla} \phi)
  \boldsymbol{t}_{w} \vert}  .
\end{equation}
Thus, in Geom one has,
\begin{equation}
  \label{eq:bc-op-geom}
 \frac{\partial \phi}{\partial n_{w}} \bigg\vert _{S} 
= - \tan \bigg( \frac{\pi}{2}-\theta_{w} \bigg)  
\vert  \boldsymbol{t}_{w} \cdot \boldsymbol{\nabla} \phi \vert  .
\end{equation}
In the design of Geom, the following fact has been taken into account: 
the tangential component of $\phi$'s gradient cannot be modified
during simulation and the local (microscopic) contact angle can only
be enforced through the change of the normal
component~\cite{pre07-geom-wbc}.
Therefore, it does better than the SE formulation
to make sure that the local contact angle matches the specified value
(which will be confirmed by the numerical results later).

As noted before, the CI-based WBC is much like Geom
except the way to enforce the microscopic contact angle.
Because the key of CI is embedded in the specific implementation
after spatial discretization, we will introduce it later in Section \ref{ssec:implem-wbcs}.

\subsection{Governing Equations and Numerical Method}

In the above, the basics of PF model for binary fluids and
five WBCs were introduced. 
Next, the governing equations and the method for their numerical solution are described.
For flows of binary fluids,
there are two types of dynamics: the hydrodynamics for fluid flow and
the interfacial dynamics.
The equation for the latter has been given, 
i.e., Eq. (\ref{eq:che}) supplemented with Eq. (\ref{eq:chem-potential}).
For axisymmetric problems, they read,
\begin{equation}
\label{eq:che-axisym}
\frac{\partial \phi}{\partial t} + u_{r} \frac{\partial \phi}{\partial r} 
+ u_{z} \frac{\partial \phi}{\partial z}  
= M  \bigg( \frac{\partial^{2} \mu}{\partial r^{2}} 
+ \frac{1}{r} \frac{\partial \mu}{\partial r} 
+ \frac{\partial^{2} \mu}{\partial z^{2}} \bigg), 
\end{equation}
\begin{equation}
  \label{eq:chem-potential-axisym}
\mu = 4 a \phi (\phi ^2 -1) 
-  \kappa  \bigg( \frac{\partial^{2} \phi}{\partial r^{2}} 
+ \frac{1}{r} \frac{\partial \phi}{\partial r} 
+ \frac{\partial^{2} \phi}{\partial z^{2}} \bigg).
\end{equation}
With the interfacial tension effects modeled by the PF model,
the governing equations for the incompressible axisymmetric flow of binary fluids
having uniform density and viscosity 
may be written as,
\begin{equation}
  \label{eq:INSCHE_continuity-cylind-axisymm-noazi}
\frac{\partial u_{r}}{\partial r} + \frac{u_{r}}{r} 
+ \frac{\partial u_{z}}{\partial z} = 0 ,
\end{equation}
\begin{equation}
  \label{eq:INSCHE_momentum-cylind-r-axisymm-noazi}
\begin{split}
&  \frac{\partial u_{r}}{\partial t} 
  + \bigg( u_{r} \frac{\partial u_{r} }{\partial r} 
+ u_{z} \frac{\partial u_{r} }{\partial z} 
\bigg)
= - \frac{\partial S_{p} }{\partial r}\\
&+ \nu \bigg( \frac{\partial^{2} u_{r}}{\partial r^{2}} 
+ \frac{1}{r} \frac{\partial u_{r}}{\partial r} 
+ \frac{\partial^{2} u_{r}}{\partial z^{2}} 
 -  \frac{u_{r}}{r^{2}}\bigg)
- \phi \frac{\partial \mu}{\partial r} ,
\end{split}
\end{equation}
\begin{equation}
  \label{eq:INSCHE_momentum-cylind-z-axisymm-noazi}
 \begin{split}
& \frac{\partial u_{z}}{\partial t} 
+ \bigg( u_{r} \frac{\partial u_{z}}{\partial r} 
+ u_{z} \frac{\partial u_{z}}{\partial z} \bigg)
= - \frac{\partial S_{p}}{\partial z}\\
&+ \nu \bigg( \frac{\partial^{2} u_{z}}{\partial r^{2}} 
+ \frac{1}{r} \frac{\partial u_{z}}{\partial r} 
+ \frac{\partial^{2} u_{z}}{\partial z^{2}} \bigg)
- \phi \frac{\partial \mu}{\partial z} ,
\end{split}
\end{equation}
where $S_{p}$ is a term similar to the hydrodynamic pressure
in single-phase incompressible flow~\cite{jacqmin99jcp}.
Note that for simplicity the density and viscosity have been assumed to be uniform
and more focus is given to the WBC. 
To use the notation for 2-D Cartesian coordinates,
we replace $(z, r)$ with $(x, y)$.
Besides, $(u_{z}, u_{r})$ are replaced by $(u, v)$.
For 2-D problems (which may be encountered in some of the following studies), 
the governing equations are simplified.
Specifically, they may be obtained by removing 
the term $M \frac{1}{r} \frac{\partial \mu}{\partial r}$
from Eq. (\ref{eq:che-axisym}),
the term $- \kappa \frac{1}{r} \frac{\partial \phi}{\partial r}$
from Eq. (\ref{eq:chem-potential-axisym}),
the term $\frac{u_{r}}{r}$  
from Eq. (\ref{eq:INSCHE_continuity-cylind-axisymm-noazi}),
the two terms, $\nu \frac{1}{r} \frac{\partial u_{r}}{\partial r} $ and 
$\nu (-  \frac{u_{r}}{r^{2}})$, 
from Eq. (\ref{eq:INSCHE_momentum-cylind-r-axisymm-noazi}),
and the term $\nu \frac{1}{r} \frac{\partial u_{z}}{\partial r}$ 
from Eq. (\ref{eq:INSCHE_momentum-cylind-z-axisymm-noazi}).

In the hybrid lattice-Boltzmann finite-difference method~\cite{hybrid-mrt-lb-fd-axisym},
the LBM is employed to simulate the hydrodynamics,
described by the Navier-Stokes equations (NSEs, specifically,
Eqs. (\ref{eq:INSCHE_continuity-cylind-axisymm-noazi}),
(\ref{eq:INSCHE_momentum-cylind-r-axisymm-noazi})
and (\ref{eq:INSCHE_momentum-cylind-z-axisymm-noazi}) for axisymmetric problems),
whereas the equation for the interface motion (the CHE, Eq. (\ref{eq:che})), 
is solved by the finite-difference method for spatial discretization 
and the $4^{th}-$order Runge-Kutta method for time marching. 
The details of this hybrid method can be found
in the Ref. ~\cite{hybrid-mrt-lb-fd-axisym} and will not be repeated.
Here more attention is paid to the specific implementations of different WBCs.
We note that there are different choices for several components
in the hybrid method in~\cite{hybrid-mrt-lb-fd-axisym}.
The present work uses the multiple-relaxation-time (MRT) collision model for LBM,
the centered formulation for the forcing terms, and
the isotropic discretization based on D2Q9 velocity model 
(i.e., the \texttt{iso} scheme in ~\cite{hybrid-mrt-lb-fd-axisym}) 
to evaluate the spatial gradients of the PF variables.

The domain of simulation is a rectangle specified by
$0 \leq x \leq L_{x}, \  0 \leq y \leq L_{y}$.
It is discretized into 
$N_{x} \times N_{y}$ uniform squares of side length $h$,
thus, $L_{x} = N_{x} h$ and $L_{y} = N_{y} h$.
The distribution functions in LBM and the discrete phase-field variables, $\phi_{i,j}$ and $\mu_{i,j}$,
are both located at the centers of the squares
(like the cell centers in the finite-volume method).
The indices $(i,j)$ for the bulk region (i.e., within the computational domain)
are $1\leq i \leq N_{x}, \ 1\leq j \leq N_{y}$.

\subsection{Implementation of Different WBCs}\label{ssec:implem-wbcs}

As seen in Section \ref{ssec:wbc}, all WBCs (except CI) involve the enforcement
of the normal gradient of the order parameter $\phi$ on the wall.
Consider the case with the lower side of a rectangle 
being a wall with a given contact angle $\theta_{w}$.
The enforcement of $\phi$'s normal gradient is realized by adding a ghost layer of squares
having the same size as those in the bulk region,
the centers of which are $h/2$ below the wall with the index $j=0$.
Although the normal gradient of $\phi$ is not directly enforced in CI,
it also specifies the value of $\phi$ in the ghost layer.

When LinSE is used, after discretization, Eq. (\ref{eq:bc-op-LinSE-simplified}) becomes,
\begin{equation}
\frac{\phi_{i,1} - \phi_{i,0}}{h} = - \frac{2}{W} \tilde{\omega},
\end{equation}
giving the order parameter in the ghost layer,
\begin{equation}
\label{eq:phi0-LinSE}
\phi_{i,0} = \phi_{i,1} + \frac{2}{W/h} \tilde{\omega} 
= \phi_{i,1} + \frac{2}{\tilde{W}} \tilde{\omega}  ,
\end{equation}
where $\tilde{W} = W/h$ is the dimensionless interface width 
(i.e., $W$ measured in the grid size $h$).
Similarly, when CubSE is used, one finds,
\begin{equation}
\label{eq:phi0-CubSE}
\phi_{i,0} = \phi_{i,1} + \frac{2}{\tilde{W}} \cos \theta_{w} (1 - \phi_{S}^{2}) ,
\end{equation}
and for SinSE one has,
\begin{equation}
\label{eq:phi0-SinSE}
\phi_{i,0} = \phi_{i,1} + \frac{2}{\tilde{W}} \cos \theta_{w} 
\bigg[ \frac{\pi}{3} \cos \bigg(  \frac{\pi}{2} \phi_{S} \bigg) \bigg] .
\end{equation}
When Geom is used, one has,
\begin{equation}
\label{eq:phi0-geom}
\phi_{i,0} = \phi_{i,1} + \tan \bigg( \frac{\pi}{2}-\theta_{w} \bigg)
\vert  \boldsymbol{t}_{w} \cdot \boldsymbol{\nabla} \phi \vert  h.
\end{equation}
It is seen that, unlike Eq. (\ref{eq:phi0-LinSE}),
Eqs. (\ref{eq:phi0-CubSE}),  (\ref{eq:phi0-SinSE}),  
and (\ref{eq:phi0-geom}) contain additional unknowns
which must be found by some means.
In Eq. (\ref{eq:phi0-CubSE}) and  (\ref{eq:phi0-SinSE}),
$\phi_{S}$ is the order parameter on the wall.
For $\theta_{w} \neq 90^{\circ}$ ($\cos \theta_{w} \neq 0$), 
when CubSE is used, 
$\phi_{S}$ can be found from the following equation (derived by assuming a quadratic profile
for $\phi$ as a function of the coordinate normal to the wall; see~\cite{jast-hybrid-lbm-fvm}
for more details),
\begin{equation}
\label{eq:phiS-CubSE}
  \phi_{S}^{2} + \frac{8}{3 q} \phi_{S} - 1 + \frac{8 \phi_{i,1}
- (\phi_{i,2} - \phi_{i,1})}{- 3 q} = 0  , \quad \textrm{with} \quad
q = \frac{2}{\tilde{W}} \cos \theta_{w}   ,
\end{equation}
and when SinSE is used, 
$\phi_{S}$ satisfies,
\begin{equation}
\label{eq:phiS-SinSE}
  8 \phi_{S} - q \pi \cos \bigg(  \frac{\pi}{2} \phi_{S} \bigg)  - (9 \phi_{i,1} - \phi_{i,2}) = 0 .
\end{equation}
Note that Eq. (\ref{eq:phiS-CubSE}) has two solutions and it is necessary to discard one
of them that is not suitable (e.g., out of the range of $\phi$).
The solution of Eq. (\ref{eq:phiS-SinSE}) may be found by Newton's method
with the initial guess of $\phi_{S}$ obtained from a simple linear extrapolation
as $\phi_{S} = 1.5 \phi_{i,1} - 0.5 \phi_{i,2}$.
For $\theta_{w} = 90^{\circ}$ ($\cos \theta_{w} = 0$), 
it is straightforward to find
from Eq. (\ref{eq:phi0-CubSE}) or Eq. (\ref{eq:phiS-SinSE})
that $\phi_{i,0} = \phi_{i,1}$.
For Geom,
Eq. (\ref{eq:phi0-geom}) contains the tangential component of $\phi$'s gradient on the wall,
$\boldsymbol{t}_{w} \cdot \boldsymbol{\nabla} \phi \vert_{S}$, 
and it is evaluated by the following extrapolation scheme,
\begin{equation}
\label{eq:tang-grad-extrap}
\boldsymbol{t}_{w} \cdot \boldsymbol{\nabla} \phi \vert_{S}
= 1.5 \boldsymbol{t}_{w} \cdot \boldsymbol{\nabla} \phi \vert_{i,1} 
- 0.5 \boldsymbol{t}_{w} \cdot \boldsymbol{\nabla} \phi \vert_{i,2}  ,
\end{equation}
where the tangential gradients on the RHS
are calculated by the $2^{nd}$-order central difference scheme, for example,
\begin{equation}
\boldsymbol{t}_{w} \cdot \boldsymbol{\nabla} \phi \vert_{i,1} 
= \frac{\partial \phi}{\partial t_{w}} \bigg\vert_{i,1} 
= \frac{\phi_{i+1,1} - \phi_{i-1,1}}{2 h}  .
\end{equation}
It is noted that in~\cite{pre07-geom-wbc}
Eq. (\ref{eq:phi0-geom}) was applied only 
in the interfacial region (specified by $0.001 < C_{i,1} < 0.999$).
Since away from the interface the gradient of the order parameter is negligible,
it is acceptable to apply Eq. (\ref{eq:phi0-geom}) everywhere along
the layer near the wall
(which is adopted in this work).

In the CI-based WBC, the values of $\phi$ in the ghost layer are found by drawing the respective characteristic lines
(contours of $\phi$) from the centers of the squares in the ghost layer, which intersect the wall at the given
contact angle $\theta_{w}$, and linearly interpolating the values of $\phi$ at the intersection points of 
such characteristic lines and the layer nearest to the wall (i.e., the layer with the index $j=1$).
The illustration of the idea may be found in Fig. 4 of~\cite{cf11-accur-cabc}
and the specific formulas to obtain $\phi_{i,0}$ were given in~\cite{cf11-accur-cabc} as well.
Both Geom and CI aim to enforce exactly 
the microscopic contact angle right on the wall directly 
(of course, with certain approximations embedded in the numerical evaluations
of derivatives or in the extrapolations and interpolations).
Thus, it is expected that they would likely give similar results.

Once the order parameter in the ghost layer below the wall is specified
according to the respective formulas, the WBC is implemented completely.
For a wall along some other directions, the formulas for different WBCs are similar
(only some changes to the indices are required).
For conciseness they are not given here.

\subsection{Inclusion of an Additional Procedure Mimicking the Wall Energy Relaxation}\label{ssec:implem-wbc-wall-relax}

In~\cite{pof09-dyn-wetting, pof11-wall-energy-relax},
an additional parameter was introduced to control the speed of the establishment of local equilibrium on the wall.
As pointed by Yue and Feng~\cite{pof11-wall-energy-relax},
the additional parameter may be explored as a phenomenological parameter to 
match the simulation results with experimental ones.
Here we propose the inclusion of an additional simple procedure in the implementation of the WBC
which serves a similar purpose.
This additional step is also partly inspired by the implementation of the 
contact angle hysteresis  (CAH) model by Ding and Spelt~\cite{jfm08-drop-shear}.
Specifically, this step is about the update of the value of $\phi$ in the ghost layer.
We still use the lower wall to illustrate the details.
In the above, the equations to obtain $\phi_{i,0}$ have been presented
(for different WBCs the equation differs, 
for instance, Eq. (\ref{eq:phi0-CubSE}) is from CubSE).
For concreteness, we take CubSE as an example:
Eq. (\ref{eq:phi0-CubSE}) is employed to update $\phi_{i,0}$ at every time step (or sub-step),
and the values of $\phi_{i,0}$ in previous steps are not used at all.
By contrast, for the CAH model implemented according to~\cite{jfm08-drop-shear}, 
when the local contact angle is between
the advancing and receding contact angles, 
$\phi_{i,0}$ is not updated
(i.e., keeps its value in previous step) 
so that the hysteresis effects are taken into account.
For convenience, denote the value of $\phi_{i,0}$ at the new step (or sub-step)
in case of no slip (i.e., completely hysteretic) 
as $\phi_{i,0}^{hy}$ (equal to its value in previous step or sub-step)
and that in case of full slip as $\phi_{i,0}^{sl}$ (its value given by Eq. (\ref{eq:phi0-CubSE})).
Now we propose the following procedure to calculate $\phi_{i,0}$ at each new time step (or sub-step),
\begin{equation}
\label{eq:phi0-wall-relax}
\phi_{i,0} = \phi_{i,0}^{hy} + r_{wr} (\phi_{i,0}^{sl} - \phi_{i,0}^{hy})
=  r_{wr} \phi_{i,0}^{sl} + (1 - r_{wr} ) \phi_{i,0}^{hy},
\end{equation}
where $r_{wr}$ ($0 \leq r_{wr} \leq 1$) 
is a newly introduced parameter to control the rate of relaxation on the wall,
and $\Delta \phi_{i,0} = \phi_{i,0}^{sl} - \phi_{i,0}^{hy}$ corresponds to the change of $\phi_{i,0}$
in case of full slip.
It is easy to see that $r_{wr} = 0$ corresponds to the case of no slip
and $r_{wr} = 1$ corresponds to the case of full slip.
As compared to the relevant equations in~\cite{pof09-dyn-wetting} and~\cite{pof11-wall-energy-relax},
Eq. (\ref{eq:phi0-wall-relax}) is more simple and 
does not involve the discretization of time derivatives.
Besides, the meaning of the new parameter $r_{wr}$ is obvious 
(i.e., the weight to control how much slip is allowed).

\section{Results and Discussions}\label{sec:res-dis}

\subsection{Characteristic Quantities, 
Dimensionless Numbers and Common Setups}
\label{ssec:char-quant-diml-num-setup}

Before presenting the results,
we first introduce the characteristic quantities and dimensionless numbers,
as well as some common setups for the problems to be studied.
In each problem, we study a drop with a radius $R$ 
or a liquid column with a height or diameter $H$, 
and $R$ or $H$ is chosen to be the characteristic length $L_{c}$
(for convenience, we use $R$ in the following general formulas).
The (constant) density is selected as the characteristic density $\rho_{c}$.
The interfacial tension is $\sigma$ and the kinematic viscosity is $\nu$ (thus,
the dynamic viscosity is $\eta = \rho_{c} \nu$).
As in some previous works~\cite{DIMSpreading07, hybrid-mrt-lb-fd-axisym}, 
here we use the following characteristic velocity $U_{c}$,
\begin{equation}
\label{eq:Uc}
  U_{c} = \frac{\sigma}{\rho_{c} \nu}   ,
\end{equation}
leading to a characteristic time $T_{c}$ given by,
\begin{equation}
\label{eq:Tc}
  T_{c} = \frac{L_{c}}{U_{c}} = \frac{R \rho_{c} \nu}{\sigma}  .
\end{equation}
For problems on drop motion, one may derive another set of characteristic quantities
(which are typically used in inviscid dynamics~\cite{jfm03-drop-coal}),
\begin{equation}
\label{eq:Uc-Tc-inv}
  U_{c, \textrm{inv}} = \sqrt{\frac{\sigma}{\rho_{c} R}}   ,
\quad
  T_{c, \textrm{inv}} = \frac{L_{c}}{U_{c, \textrm{inv}}} 
  = \frac{R}{U_{c, \textrm{inv}}}  = \sqrt{\frac{\rho_{c} R^{3}}{\sigma}}  .
\end{equation}
All other quantities of length,
time and velocity below are scaled by $L_{c}$, $T_{c}$ and $U_{c}$ by default
(sometimes $T_{c, \textrm{inv}}$ and $U_{c, \textrm{inv}}$ are used instead of $T_{c}$ and $U_{c}$).
There are two important physical parameters in drop problems:
(1) the capillary number, which reflects the ratio of the viscous force
over the interfacial tension force, and with the above definition of $U_{c}$,
is found to be always unity,
\begin{equation}
\label{eq:ca-def}
  Ca = \frac{\rho_{c} \nu U_{c}}{\sigma} = 1  ,
\end{equation}
(2) the Reynolds number, which reflects the ratio of the inertial force
over the viscous force and is found to be (if $U_{c}$ is used), 
\begin{equation}
\label{eq:re-def}
  Re = \frac{U_{c} R}{\nu} = \frac{\sigma}{\rho_{c} \nu}
\frac{R}{\nu} = \frac{\sigma R}{\rho_{c} \nu^{2}}  .
\end{equation}
It is noted that based on $U_{c, \textrm{inv}}$ one may define another 
capillary number and Reynolds number as~\cite{jfm05drop-coal},
\begin{equation}
\label{eq:ca-re-inv-def}
  Ca_{\sigma} = \frac{\rho_{c} \nu U_{c, \textrm{inv}}}{\sigma} = \frac{1}{\sqrt{Re}}  ,
\quad
  Re_{\sigma} = \frac{U_{c, \textrm{inv}} R}{\nu} =   \sqrt{Re} .
\end{equation}
In addition, the Ohnesorge number $Oh$
is also often used for drop dynamics~\cite{arfm06-drop-impact}.
If the drop radius (instead of the drop diameter in~\cite{arfm06-drop-impact}) is used,
$Oh$ reads,
\begin{equation}
\label{eq:oh-def}
  Oh = \frac{\rho_{c} \nu}{\sqrt{\rho_{c} \sigma R}} ,
\end{equation}
which is found to be related to the other dimensionless numbers as 
$Oh = 1 / \sqrt{Re} = 1 / Re_{\sigma}  = Ca_{\sigma} $.

In PF simulations, there are two additional parameters: (1) the Cahn number, 
defined to be the ratio of interface width
over the characteristic length,
\begin{equation}
Cn = \frac{W}{L_{c}}  ,
\end{equation}
(2) the Peclet number,  measuring the relative magnitude of convection
over diffusion in the CHE,
\begin{equation}
Pe = \frac{U_{c} L_{c}^{2}}{M \sigma}  .
\end{equation}
We note that Yue et al.~\cite{jfm10-sil-che-cl} studied the convergence of numerical results
by PF simulations towards the sharp-interface limit and proposed the use of 
an alternative parameter $S$ (instead of $Pe$) defined by,
\begin{equation}
  S = \frac{\sqrt{M \nu}}{L_{c}},
\end{equation}
which reflects the diffusion length scale at the contact line
(relative to the characteristic length)~\cite{jfm10-sil-che-cl}
and could be more appropriate for problems involving contact lines.
In this work both $Pe$ and $S$ are provided.
In the literature, there are some investigations and discussions
on how to choose $Cn$ and $Pe$ (or $S$) to obtain reliable results
for different problems~\cite{jacqmin99jcp, jfm10-sil-che-cl}. 
We will also carry out some studies in this aspect
for one of the problems  below (drop dewetting).
It is noted that the present definition of Cahn number differs from some others
because of different definitions of the interface width.
In~\cite{jfm10-sil-che-cl, jcp07-dim-ldr}
the interface width $\varepsilon$ is related to the present one as 
$\varepsilon = W / (2 \sqrt{2})$
whereas in~\cite{ijmf06-cap-dri-flows} the interface width $\xi$ is related to $W$ as
$\xi = W / \sqrt{2}$.

In all simulations, the lower side is the symmetric axis
on which the symmetric boundary conditions are applied,
and the upper side is a stationary solid wall with wall boundary conditions applied.
The wettability of the upper wall and
the boundary types of the left and right sides differ in different problems, 
and they will be stated later individually.
The simulations are performed in the temporal range
$0 \leq t \leq t_{e}$,
where $t_{e}$ denotes the time (measured in $T_{c}$ by default, or in $T_{c, \textrm{inv}}$ when specified so)
at the end of the simulation.
Suppose the characteristic length $L_{c}$ 
is discretized by $N_{L}$ uniform segments
and the characteristic time $T_{c}$ 
(as defined in Eq. (\ref{eq:Tc}))
is discretized by $N_{t}$ uniform segments,
then one has the grid size $h$ and time step $\delta_{t}$ as follows,
\begin{equation}
  h= \frac{L_{c}}{N_{L}} ,
\quad \delta_{t} = \frac{T_{c}}{N_{t}}  .
\end{equation}
Recall that the domain of size $L_{x} \times L_{y}$
is discretized into $N_{x} \times N_{y}$ uniform squares
($L_{x} = N_{x} h$ and $L_{y} = N_{y} h$),
one has $h = L_{c} / N_{L} = L_{x} / N_{x} = L_{y} / N_{y}$.

\subsection{Common Quantities of Interest}

\begin{figure}[htp]
  \centering
  \includegraphics[scale = 0.5]{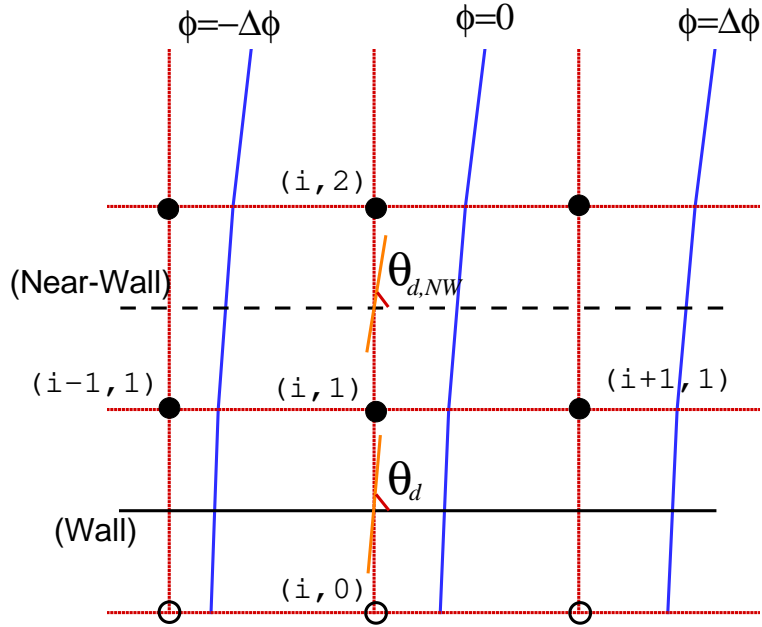}
  \caption{Illustration of the definitions and calculations of 
    the local contact angle on the wall, $\theta_{d}$,
    and that near the wall, $\theta_{d, NW}$.}
  \label{fig:dca-cal}
\end{figure}

In all problems, we are concerned about 
the centroid (average) velocity (of the drop or liquid column) 
in the $x$-direction. 
Take the drop under the cylindrical geometry as an example:
the average velocity $\overline{U}_{d} (t)$ may be calculated by, 
\begin{equation}
\label{eq:drop-U}
      \overline{U}_{d} (t)
      = \frac{\int_{V} N (\phi) u (t) d \Omega}{\int_{V} N (\phi) d \Omega} 
      \approx \frac{\sum_{i,j} y_{i,j} u_{i,j} (t) N (\phi_{i,j})}{
        \sum_{i,j} y_{i,j} N (\phi_{i,j})}   ,
\end{equation}
where $V$ denotes the domain, 
$d \Omega (= 2 \pi r dr dz) = 2 \pi y dy dx \approx (2 \pi h^{2}) y_{i,j} $
under the cylindrical geometry,
and the function $N (\phi)$ is defined by,
\begin{equation}
  N (\phi) = \left \{
\begin{array}{cl}
1 & \quad \textrm{if} \quad \phi > 0 \\
0 & \quad \textrm{if} \quad \phi \leq 0 
\end{array} \right .  .
\end{equation}
For a liquid column under the same geometry, 
the centroid (average) velocity (denoted by $v_{\textrm{lc}}$)
is also calculated by Eq. (\ref{eq:drop-U}) 
(under 2-D geometry, the formula is more simple because $d \Omega =  dx dy$).
In addition, from Eq. (\ref{eq:geom-ca}), a \textit{local} dynamic contact angle 
$\theta_{d, l}$ (in degree $^{\circ}$) 
may be calculated from the local gradients of the order parameter 
\textit{on the wall} as,
\begin{equation}
  \label{eq:LocalCA}
  \theta_{d, l} = \frac{180^{\circ}}{\pi} \bigg( \frac{\pi}{2}
   - \arctan \frac{ - \frac{\partial \phi}{\partial n_{w}}}{\vert \frac{\partial \phi}{\partial t_{w}} \vert}  \bigg) .
\end{equation}
It should be noted that far away from the interfaces
$\vert \frac{\partial \phi}{\partial t_{w}} \vert$ may become zero
and Eq. (\ref{eq:geom-ca}) is thus for the interfacial region only.
In the results presented below,
the interfacial region is specified by the following conditions:
$\vert \frac{\partial \phi}{\partial t_{w}} \vert * h > 0.1$ 
and $-0.998 \leq \phi \leq 0.998$;
outside this region, we set $\theta_{d, l} = 0$.
Near the interface (where $\phi=0$) the dynamic contact angle 
on the wall is denoted simply as $\theta_{d}$.
Besides, the dynamic contact angle \textit{near} the wall
(obtained one grid away from the wall) 
is denoted as $\theta_{d, NW}$.
These two dynamic contact angles 
and the nodes involved in their calculation
are illustrated in Fig. \ref{fig:dca-cal}
for the case with a wall on the lower side. 

\subsection{Study of the Static Contact Angle}\label{ssec:staticca}

\begin{figure}[htp]
  \centering
  \includegraphics[scale = 0.58]{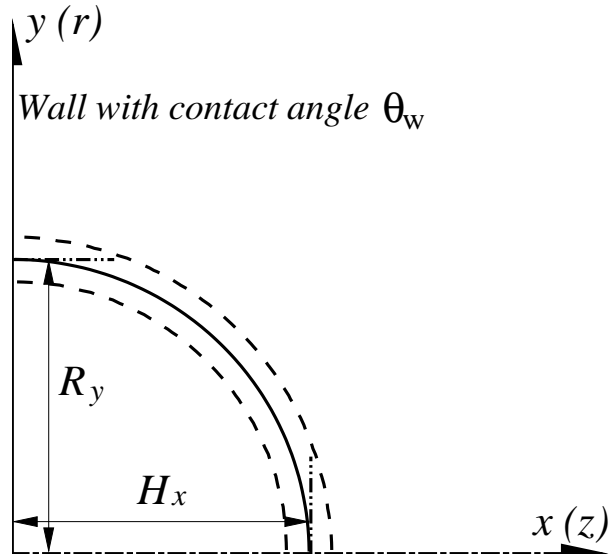}
  \caption{Illustration of the problem setup for the study of static contact angle 
  (and also for drop dewetting). 
  The outer and inner dashed lines
  correspond to $\phi = - 0.9$ and $\phi = 0.9$, respectively,
  whereas the solid line corresponds to $\phi = 0$.}
  \label{fig:problem-setup}
\end{figure}

First, we study the static contact angle of a drop on a homogeneous wall.
Specifically, we investigate whether a drop having an initial shape corresponding 
to an initial contact angle of $\theta_{i}$ 
and also in touch of a wall with the wettability specified by an contact angle $\theta_{w}$
($\theta_{w} \neq \theta_{i}$) can evolve to
reach an equilibrium state with a contact angle equal to $\theta_{w}$.
In this study (and that on drop dewetting), 
the other common setups besides those given above are as follows.
Both the left and right sides are stationary walls.
The contact angle of the left wall is $\theta_{w}$ which may take different values,
and those of the upper and right walls are 
$90^{\circ}$ (which is not quite important because the drop does not touch them).
The initial position of the drop center 
(actually the center of the circle of which the middle cross section of the drop is a part)
is $(x_{c}, y_{c})$
with $y_{c} = 0$ (as required by the fact that the lower side is a line of symmetry).
From simple geometrical relations, the initial (given) contact angle 
$\theta_{i}$ (in degree $^{\circ}$)
corresponding to the initial drop shape is related to $x_{c}$ as,
\begin{equation}
\label{eq:thetai}
  \theta_{i} = \frac{180^{\circ}}{\pi} \times \left \{ \begin{array}{cl}
   (\pi - \arccos \frac{x_{c}}{R}) &  \quad \textrm{if}  \quad x_{c} > 0  \\
    \arccos \frac{-x_{c}}{R} &  \quad \textrm{if}  \quad x_{c} \leq 0
    \end{array} \right .  .
\end{equation}
In this work, we fix the initial drop center as $(x_{c}, y_{c}) = (0,0)$, giving 
an initial contact angle $\theta_{i} = 90^{\circ}$.
The domain size is $L_{x} \times L_{y} = 4 \times 4$.
Figure \ref{fig:problem-setup} illustrates the basic setup 
and also the initial condition for the study of static contact angle (and also for drop dewetting).
Note that only the region near the drop is shown,
and the upper and right walls are not included in Fig. \ref{fig:problem-setup}.
We monitor
the drop \emph{height} on the $x-$axis $H_{x} = H_{x} (t)$
and the drop \textit{radius} on the left wall $R_{y} = R_{y} (t)$
(i.e., the radius of the circle on the wall formed by the contact line). 
The illustrations of $H_{x}$ and $R_{y}$ are also given in Fig. \ref{fig:problem-setup}.
If the drop shape (in the middle cross section) is part of a circle 
(which is reasonable if inertial effects are relatively small or negligible), 
the instantaneous contact angle $\theta_{d, \textrm{sf}} = \theta_{d, \textrm{sf}} (t)$ 
(obtained by fitting the drop shape at $t$, 
also in degree $^{\circ}$) may be deduced 
from $H_{x}$ and $R_{y}$ as (using the relation similar to Eq. (\ref{eq:thetai}))
~\cite{ijnmf-bubble-entrap},
\begin{equation}
  \label{eq:CA-hy-hx}
  \theta_{d, \textrm{sf}} = \frac{180^{\circ}}{\pi} 
\bigg(\pi - \arccos \bigg( 
\frac{1 - k_{r}^{2}}{1 + k_{r}^{2}} \bigg) \bigg)  ,
\quad \textrm{with} 
\quad k_{r} = R_{y} / H_{x}.
\end{equation}
Even if the inertial effect plays a role during the evolution,
the above assumption for calculating $\theta_{d, \textrm{sf}}$ is well satisfied at the end ($t = t_{e}$)
when the static equilibrium is almost reached.

\begin{figure}[htp]
  \centering
  \includegraphics[scale = 0.8]{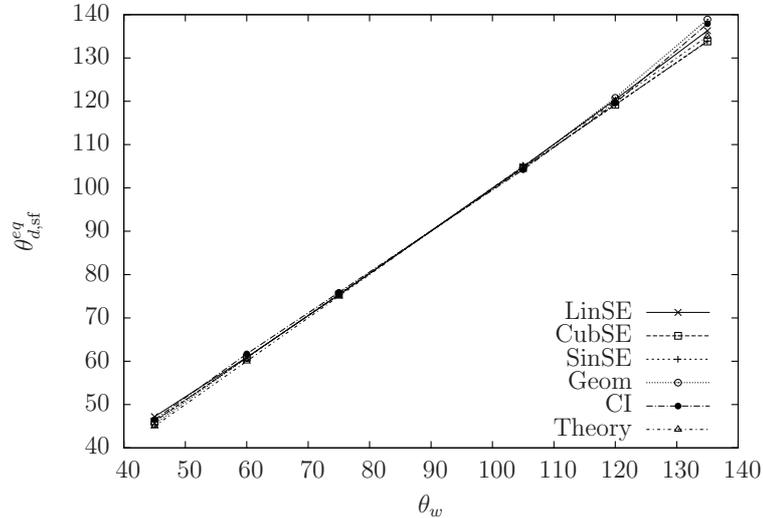}
  \caption{Comparison of the equilibrium contact angle $\theta_{d, \textrm{sf}}^{eq}$ 
  at different values of contact angle $\theta_{w}$ 
  by using different WBCs.
  The parameters are $Re=1000$ ($Oh = 0.032$), $Ca=1$,
  $Cn=0.2$, $Pe=5000$ ($S = 0.014$), $N_{L} = 20$, $N_{t} = 80$, $t_{e} = 500$.}
  \label{fig:cmp-ca-Re1k-cn0d2-Pe5k}
\end{figure}

For this study, the physical parameters are $Re = 1000$ ($Oh = 0.032$), $Ca=1$, 
and the numerical parameters are
$Cn = 0.2$, $Pe = 5000$ ($S = 0.014$),
$N_{L} = 20$ ($\tilde{W} = 4$), $N_{t} = 80$, $t_{e} =500$ 
(large enough to ensure that $\theta_{d, \textrm{sf}}$ becomes almost constant).
The focus of this problem is the final state and the temporal evolution is not concerned.
Figure \ref{fig:cmp-ca-Re1k-cn0d2-Pe5k} compares 
the equilibrium contact angle $\theta_{d, \textrm{sf}}^{eq}$ 
(as obtained from $H_{x}$ and $R_{y}$ using Eq. (\ref{eq:CA-hy-hx}) at $t=500$) 
under several given contact angles ($\theta_{w}=45^{\circ}$, 
$60^{\circ}$, $75^{\circ}$, $105^{\circ}$, $120^{\circ}$, and $135^{\circ}$)
of the left wall by using LinSE, CubSE, SinSE, Geom and CI.
It is seen that for each $\theta_{w}$
upon reaching equilibrium all the WBCs can give a contact angle
that agrees reasonably well with the given one. 
For $\theta_{w}$ being far away from $90^{\circ}$,
the deviations become somewhat larger, 
but the numerical ones are still fairly close to the theoretical values.
Thus, the differences between the five WBCs, if any, should mainly 
exist under dynamic situations.

\subsection{Study of a Liquid Column Driven by a Poiseuille Flow}\label{ssec:psld-drop}

\begin{figure}[htp]
  \centering
 \includegraphics[trim = 10mm 10mm 10mm 120mm, clip, width=12cm, scale = 1.0]{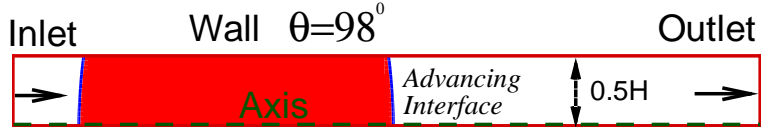}
  \caption{Problem setup for a liquid column inside a cylindrical tube 
driven by an imposed Poiseuille flow. Only the upper half of the middle cross section is shown.} 
  \label{fig:psle-drop-setup}
\end{figure}

Next, we will study some dynamic problems.
This first is the steady motion of 
a liquid column inside a cylindrical tube with a diameter $H$
driven by a Poiseuille flow.
The parabolic Poiseuille velocity profile is imposed on the left and right sides (inlet and outlet).
The average of this Poiseuille velocity profile is $\overline{V}$.
This is also an axisymmetric problem and can be simplified as a pseudo 2-D problem.
Again, the symmetry about the axis ($y = 0$) allows us to use 
only the upper half domain ($0 \leq y \leq 0.5 H$) in simulation.
The characteristic length is the tube diameter ($L_{c} = H$).
Figure \ref{fig:psle-drop-setup} illustrates the setup for this problem.
Initially, the liquid column has a width of $W_{\textrm{lc}} = 2 H$
and its center is located at $(1.5 H, 0)$.
The (static) contact angle of the tube wall (the upper wall in Fig. \ref{fig:psle-drop-setup})
is $\theta_{w} = 98^{\circ}$.
On the left and right sides, inlet and outlet boundary conditions are employed
for the distribution functions (in the ghost layers which are not involved 
in LBM calculations): 
the equilibrium parts are calculated from the given parabolic Poiseuille velocity profile
whereas the non-equilibrium parts are obtained from bounce-back rules.
For the PF variables, periodic boundary conditions are employed
on these two sides.
Driven by the imposed Poiseuille flow, the liquid column is gradually accelerated towards the right end
and achieves a steady motion after certain time.
In most of the cases studied here, the length of the domain is $L_{x} = 5 H$,
which is long enough for the liquid column to achieve the steady state
(note in some cases it is extended up to $L_{x} = 15 H$ to satisfy this requirement). 
To reach the steady state, the simulation time ranges from $3$ to $10$ $T_{c, \textrm{inv}}$ 
depending on the specific parameters like 
the average velocity $\overline{V}$ and the relaxation parameter $r_{wr}$ in the boundary condition.
For this study, the physical parameters are $Re = 100$ ($Oh = 0.1$), $Ca=1$, 
and the numerical parameters are
$Cn = 0.125$, $Pe = 5000$ ($S = 0.014$),
$N_{L} = 32$ ($\tilde{W} = 4$).
The temporal discretization parameter $N_{t}$ was varied from $320$ to $3200$ 
depending on the average velocity $\overline{V}$.

In this problem, we are 
only concerned about the steady state, in which
the liquid column moves at a constant speed 
(just equal to 
$\overline{V}$).
Based on $\overline{V}$, one can define another capillary number as,
\begin{equation}
\label{eq:psle-drop-Ca-def}
Ca_{\overline{V}} = \frac{\rho_{c} \nu_{A} \overline{V}}{\sigma} ,
\end{equation}
where $\nu_{A}$ is the kinematic viscosity of the liquid column.
We focus on the advancing interface on the right.
It is noted that there is no obvious difference in the advancing interfaces
between a liquid column of a finite length 
(which is long enough to keep the advancing and receding interfaces apart during the motion)
and that of an infinite length
when the steady state is reached.
This allows us to compare the current simulations with those reported by
Fermigier and Jenffer~\cite{jcis91-exp-dca},
who studied the motion of a liquid-liquid interface inside a tube.
Two key dimensionless parameters were mentioned in their experiments,
namely, the capillary number 
($Ca_{\overline{V}}$ defined here)
and the (dynamic) viscosity ratio $r_{\eta}$
(defined as $r_{\eta} = \eta_{A} / \eta_{B}$ with $\eta_{A}$ and $\eta_{B}$ being the dynamic viscosities 
of the displacing liquid and the displaced liquid, respectively).
Here we focus on one set of experimental data reported in their work
with a viscosity ratio $r_{\eta} = 0.9$.
For simplicity the viscosity ratio in our simulations is set to be $r_{\eta} = 1$ 
(without introducing large deviations).
The advancing contact angle $\theta_{A}$ is obtained by least-square fitting 
the middle part ($0 \leq y \leq 0.25 H$) of the advancing interface with a circle.
It was found that the velocity of the liquid column became steady relatively fast
(the time taken depends on the capillary number).
The steady velocity of the liquid column obtained numerically by using Eq. (\ref{eq:drop-U})
matched the average of the imposed Poiseuille flow ($\overline{V}$) very well
for all capillary numbers and WBCs.
At the same time, it took longer time for $\theta_{A}$ to become \textit{steady}:
in fact, it was found that $\theta_{A}$ fluctuates around some value for each $Ca_{\overline{V}}$
even after long time and the oscillation amplitude decreases with increasing $Ca_{\overline{V}}$.
Fortunately, the fluctuations remained small (around $2^{\circ}$ for the smallest $Ca_{\overline{V}}$ considered).
For convenience, the value of $\theta_{A}$ at the end of simulation 
is taken.
Besides $\theta_{A}$, we also looked into 
the dynamic contact angle on the upper wall 
(at $y = 0.5 H$ near the advancing interface where $\phi = 0$),
$\theta_{d}^{\textrm{rig}}$, 
and the dynamic contact angle \textit{near} the upper wall,
$\theta_{d, NW}^{\textrm{rig}}$,
obtained one grid away from the wall 
(at $y = 0.5 H - h$) near the interface.

\begin{figure}[htp]
  \centering
  \includegraphics[scale = 1.0]{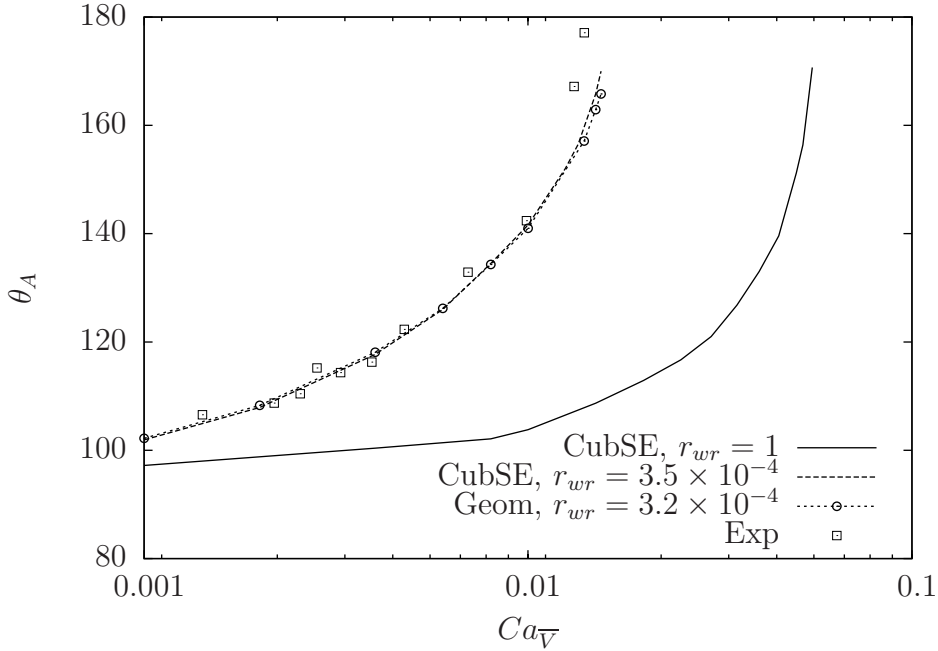}
  \caption{Variations of the advancing contact angle $\theta_{A}$
  with the capillary number $Ca_{\overline{V}}$
  by current simulations using CubSE with $r_{wr} = 3.5 \times 10^{-4}$
  and $r_{wr} = 1$ (full slip),
  using Geom with $r_{wr} = 3.2 \times 10^{-4}$,
  and also from the experiments by Fermigier and Jenffer~\cite{jcis91-exp-dca}.}
    \label{cmp-psle-drop-axisym-Ca-aca}
\end{figure}

It was observed that when the full slip condition was used (i.e. $r_{wr} = 1$),
different WBCs gave almost the same results for both $v_{\textrm{lc}}$ and $\theta_{A}$
even though the dynamic contact angles on and near the wall 
($\theta_{d}^{\textrm{rig}}$ and $\theta_{d, NW}^{\textrm{rig}}$) 
were slightly different 
(the results given by LinSE differed from the others the most).
For conciseness, the detailed results are not shown here.
In addition, the role of \textrm{wall energy relaxation} 
was also studied for two WBCs (CubSE and Geom).
Figure \ref{cmp-psle-drop-axisym-Ca-aca} shows the variations of the advancing contact angle
$\theta_{A}$ (obtained by the least-square fitting of the middle portion of the interface as mentioned above) 
with the capillary number $Ca_{\overline{V}}$
from both the present simulations with different WBCs and different values of the relaxation parameter $r_{wr}$,
and the experimental measurements by Fermigier and Jenffer~\cite{jcis91-exp-dca}.
It is seen the prediction by simulations using CubSE with $r_{wr} = 1$
differ from the experimental data significantly.
By properly adjusting the relaxation parameter $r_{wr}$,
both CubSE and Geom can give very good predictions that match closely the experimental data
over a wide range of capillary number.
It is noted that the substantial improvement due to proper wall energy relaxation 
has been reported by Yue and Feng~\cite{pof11-wall-energy-relax}
in the study of a similar problem
and by Carlson et al.~\cite{pof09-dyn-wetting}
in the study of very fast drop spreading.
For the two WBCs (CubSE and Geom), the main difference seems to be only 
in the suitable value of relaxation parameter $r_{wr}$
(which differ slightly: $3.5 \times 10^{4}$ for CubSE and $3.2 \times 10^{4}$ for Geom).
The agreement becomes not so good when $Ca_{\overline{V}}$ is so large that
$\theta_{A}$ is near $180^{\circ}$.
This may be caused by the fact that when $\theta_{A}$ is near $180^{\circ}$ 
(wetting failure is about to occur)
it is difficult to accurately fit the interface with an arc,
and large deviations exist in $\theta_{A}$ obtained by the least-square fitting.
Based on the above observations, it may be concluded that
\textit{for mechanically driven two-phase flows
it is not quite sensitive on which WBCs to be used
if the feature of bulk flow (away from the wall) is concerned}.

\subsection{Study of a Liquid Column Driven by Wettability Gradient}\label{ssec:lc-wg}

\begin{figure}[htp]
  \centering
 \includegraphics[trim = 10mm 40mm 10mm 120mm, clip, width=12cm, scale = 1.0]{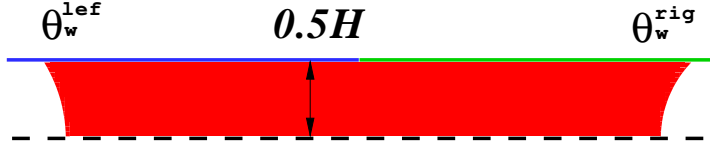}
  \caption{Problem setup for a liquid column inside a channel with a height $H$ (under 2-D geometry)
   or a cylindrical tube with a diameter $H$ (under cylinderical geometry) 
subject to a stepwise wettability gradient.}
  \label{fig:wg-rec-drop-setup}
\end{figure}

The second dynamic problem has a different driving mechanism:
in stead of an imposed flow, a given wettability gradient (WG)
is applied to drive a liquid column.
Specifically, we consider a liquid column inside a channel composed of two horizontal flat plates 
(located at $y=-0.5 H$ and $y=0.5 H$) under 2-D geometry
and also another one inside a cylindrical tube (with a diameter $H$).
The 2-D case was used as a validation case recently in~\cite{pof-sub-wgdrop-cah}
and a similar problem was investigated in~\cite{jsm12-wg-drop-lbm-sim}.
The problem under cylindrical geometry is studied here for the first time (as far as we know),
and it can also be simplified as a pseudo 2-D problem.
For completeness, we briefly reintroduce the problem setup here.
Under either 2-D or cylindrical geometry,
the problem is symmetric about the middle horizontal line $y = 0$, 
thus only the upper half ($0 \leq y \leq 0.5 H$) is used. 
The characteristic length is chosen to be the channel height or the tube diameter ($L_{c} = H$).
Figure \ref{fig:wg-rec-drop-setup} illustrates the problem setup.
Note only part of the upper half (the region near the liquid column) is shown in Fig. \ref{fig:wg-rec-drop-setup} 
because of the relatively large domain length ($L_{x} \gg H$).
Initially, the liquid column has a (nominal) width of $W_{\textrm{lc}} = 4 H$ 
(the distance between the two three-phase points (TPPs) in $x-$direction is roughly $W_{\textrm{lc}}$).
The $x-$coordinate of the middle point between the two TPPs is $x^{\textrm{mid}} = 3.5 H$,
giving the $x-$coordinates of the left and right TPPs as 
$x^{\textrm{lef}} = 1.5 H$ and $x^{\textrm{rig}} = 5.5 H$.
In the region $x> x^{\textrm{mid}}$ the wettability of the wall is 
specified by a (static) contact angle $\theta_{w}^{\textrm{rig}}$,
and for $x \leq x^{\textrm{mid}}$ the contact angle is  
$\theta_{w}^{\textrm{lef}}$ ($\theta_{w}^{\textrm{lef}} > \theta_{w}^{\textrm{rig}}$).
The initial right and left interface shapes are specified 
to be two arcs that intersect the wall with angles
$\theta_{w}^{\textrm{rig}}$ and $\theta_{w}^{\textrm{lef}}$, respectively.
Both the right and left parts of the wall are assumed to be geometrically smooth
and have no hysteresis. 
Because of the difference in the contact angle (which results in a stepwise WG), 
the liquid column is driven by the interfacial tension forces to move right
(i.e., towards the more lyophilic part). 
Boundary conditions for a stationary wall 
are applied on the upper side ($y = 0.5 H$).
Periodic boundaries are assumed on the left and right sides. 
The length of the domain is $L_{x} = 20 H$. 
To ensure that the liquid column is always under the action of the WG,
the position of the middle point $x^{\textrm{mid}} = (x^{\textrm{lef}} + x^{\textrm{rig}}) / 2$ is monitored at each step
and the wettability distribution is updated based on $x^{\textrm{mid}}$ to maintain the WG.

After certain time, the liquid column gradually reaches a steady state,
which indicates a balance between the interfacial tension forces and the viscous resistances.
Note that the dynamic contact angle on the wall $\theta_{d}$
could be equal or not equal to the static contact angle $\theta_{w}$,
depending on the WBC used.
By assuming that the velocity across the channel takes a parabolic Poiseuille velocity profile
(which should be acceptable if the regions covered by the interfaces occupy only a small portion
of the whole domain),
Esmaili et al.~\cite{jsm12-wg-drop-lbm-sim} obtained an approximate analytical solution
for the evolution of the centroid velocity of the liquid column $v_{\textrm{lc}}$ 
under 2-D geometry as
(the equation given here has been simplified for cases with uniform density $\rho_{c}$),
\begin{equation}
\label{eq:vlc-theo}
v_{\textrm{lc}} = \frac{\sigma H [2 (\cos \theta_{d}^{\textrm{rig}} - \cos \theta_{d}^{\textrm{lef}})] 
}{12 \rho_{c} [\nu_{A} W_{\textrm{lc}} + \nu_{B} (L_{x} - W_{\textrm{lc}}) ]}  
( 1 - e^{-t/t_{s}}),
\end{equation}
with $t_{s} = H^{2} L_{x} / [12 (\nu_{A} W_{\textrm{lc}} + \nu_{B} (L_{x} - W_{\textrm{lc}}) )]$
where $\nu_{B}$ is the kinematic viscosity of the fluid outside the liquid column,
and $\theta_{d}^{\textrm{rig}}$ and $\theta_{d}^{\textrm{lef}}$ are the dynamic contact angles
at the right and left TPPs
(note it is not clear whether the dynamic contact angles 
were calculated by Esmaili et al.~\cite{jsm12-wg-drop-lbm-sim}
in the same way as described above).
Here we have further derived the formula 
under cylindrical geometry
based on similar assumptions, which reads,
\begin{equation}
\label{eq:vlc-theo-axisym}
v_{\textrm{lc}} = \frac{\sigma H (\cos \theta_{d}^{\textrm{rig}} - \cos \theta_{d}^{\textrm{lef}}) 
}{8 \rho_{c} [\nu_{A} W_{\textrm{lc}} + \nu_{B} (L_{x} - W_{\textrm{lc}}) ]}  
( 1 - e^{-t/t_{s}}),
\end{equation}
with $t_{s} = H^{2} L_{x} / [32 (\nu_{A} W_{\textrm{lc}} + \nu_{B} (L_{x} - W_{\textrm{lc}}) )]$.
From the above equations, it is seen that as $t/t_{s} \rightarrow \infty$
the velocity approaches a constant value $V_{\textrm{lc}}$
for either the 2-D or axisymmetric case, specifically, 
\begin{equation}
\label{eq:Vlc-theo-2d-axisym}
V_{\textrm{lc}}^{\textrm{2D}} = \frac{\sigma H [2 (\cos \theta_{d}^{\textrm{rig}} - \cos \theta_{d}^{\textrm{lef}})] 
}{12 \rho_{c} [\nu_{A} W_{\textrm{lc}} + \nu_{B} (L_{x} - W_{\textrm{lc}}) ]}  , \quad
V_{\textrm{lc}}^{\textrm{Axisym}} = \frac{\sigma H (\cos \theta_{d}^{\textrm{rig}} - \cos \theta_{d}^{\textrm{lef}}) 
}{8 \rho_{c} [\nu_{A} W_{\textrm{lc}} + \nu_{B} (L_{x} - W_{\textrm{lc}}) ]}  .
\end{equation}
It must be noted that in the derivation of Eqs. (\ref{eq:vlc-theo}) and (\ref{eq:vlc-theo-axisym})
it is assumed that both dynamic contact angles, 
$\theta_{d}^{\textrm{rig}}$ and $\theta_{d}^{\textrm{lef}}$,
are constant.
This assumption might not always be true.
However, even if they are time-dependent, usually they should vary in a small range
for the parameters considered here.
In that case, one may express the dynamic contact angle as,
\begin{equation}
\label{eq:DCA-t}
\theta_{d} (t) = \theta_{d} (0) [1 + \epsilon(t)], \quad \textrm{where} \quad \vert \epsilon(t) \vert \ll 1.
\end{equation}
Then Eqs. (\ref{eq:vlc-theo}) and (\ref{eq:vlc-theo-axisym})
may be obtained when the leading order terms are used.

\begin{figure}[htp]
  \centering
(a)  \includegraphics[scale = 0.58]{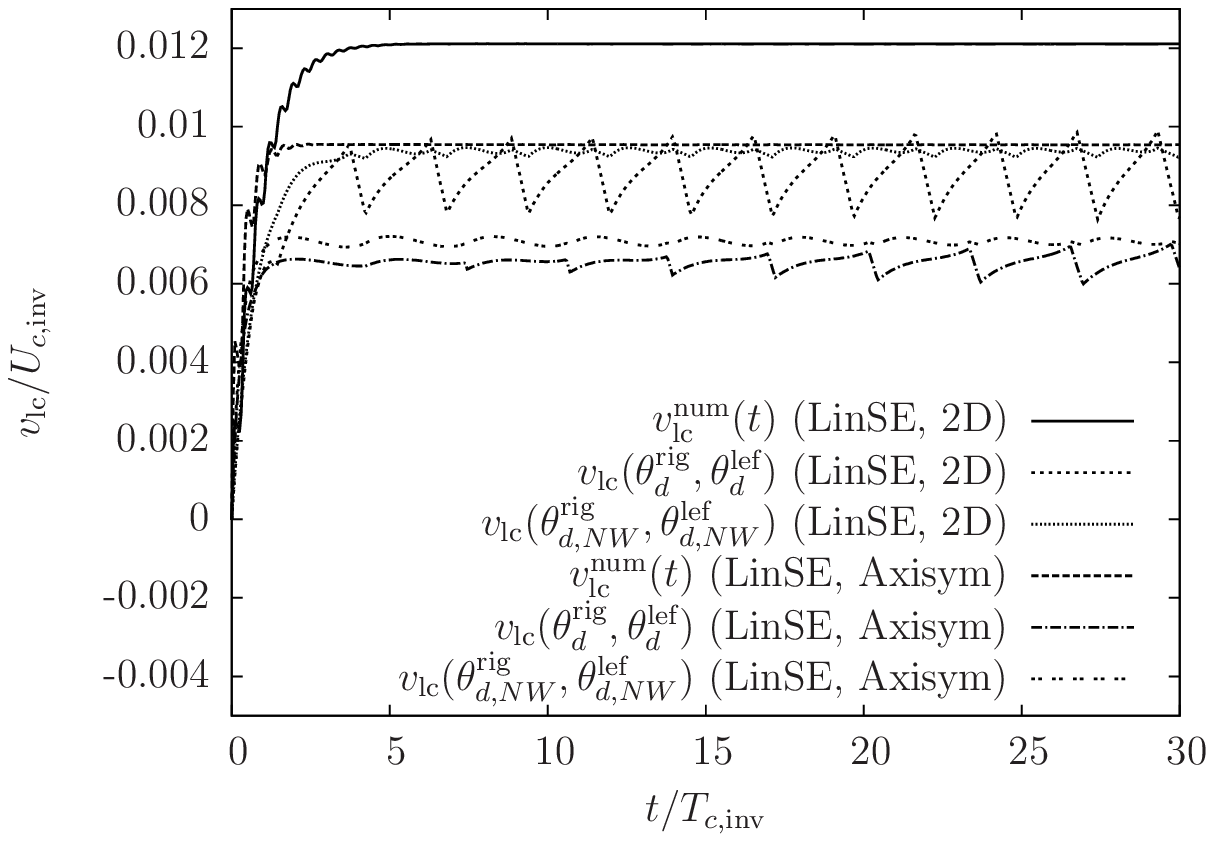}
(b)  \includegraphics[scale = 0.58]{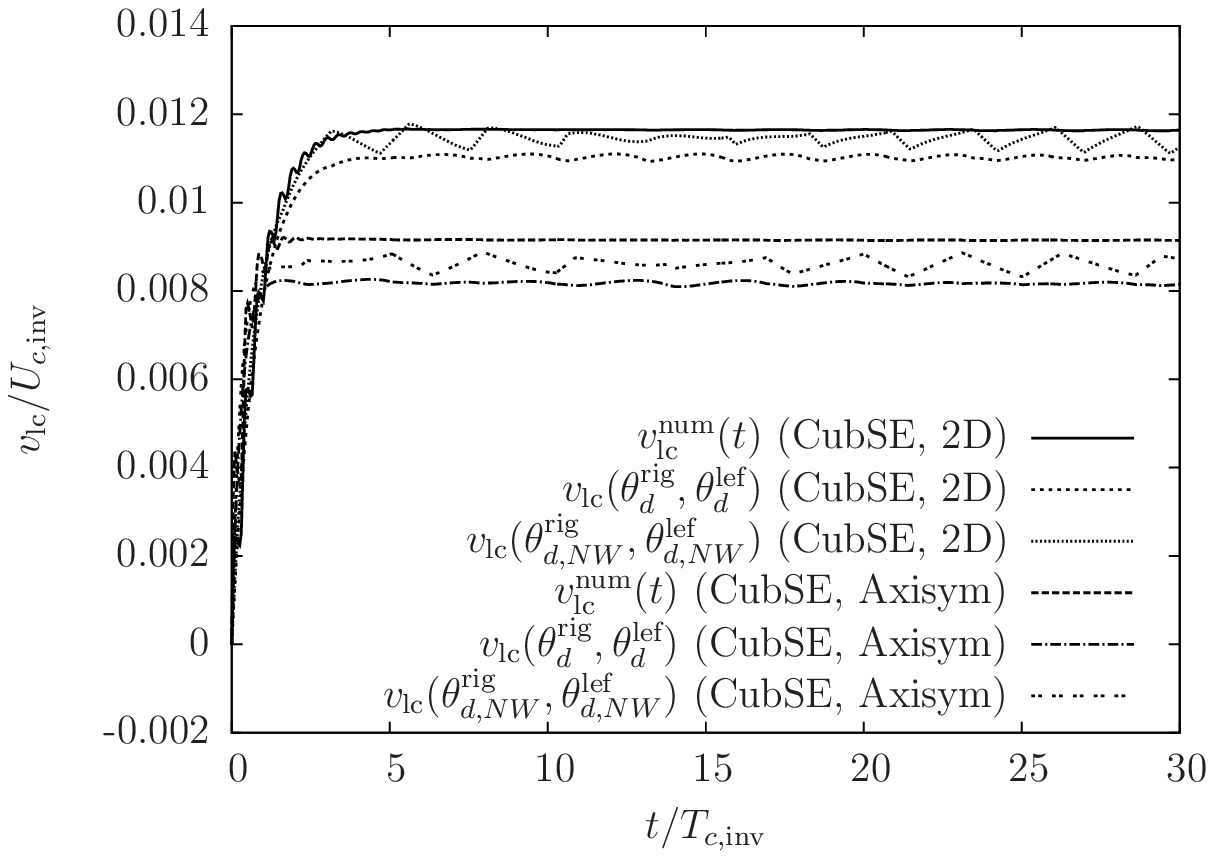}
(c)  \includegraphics[scale = 0.58]{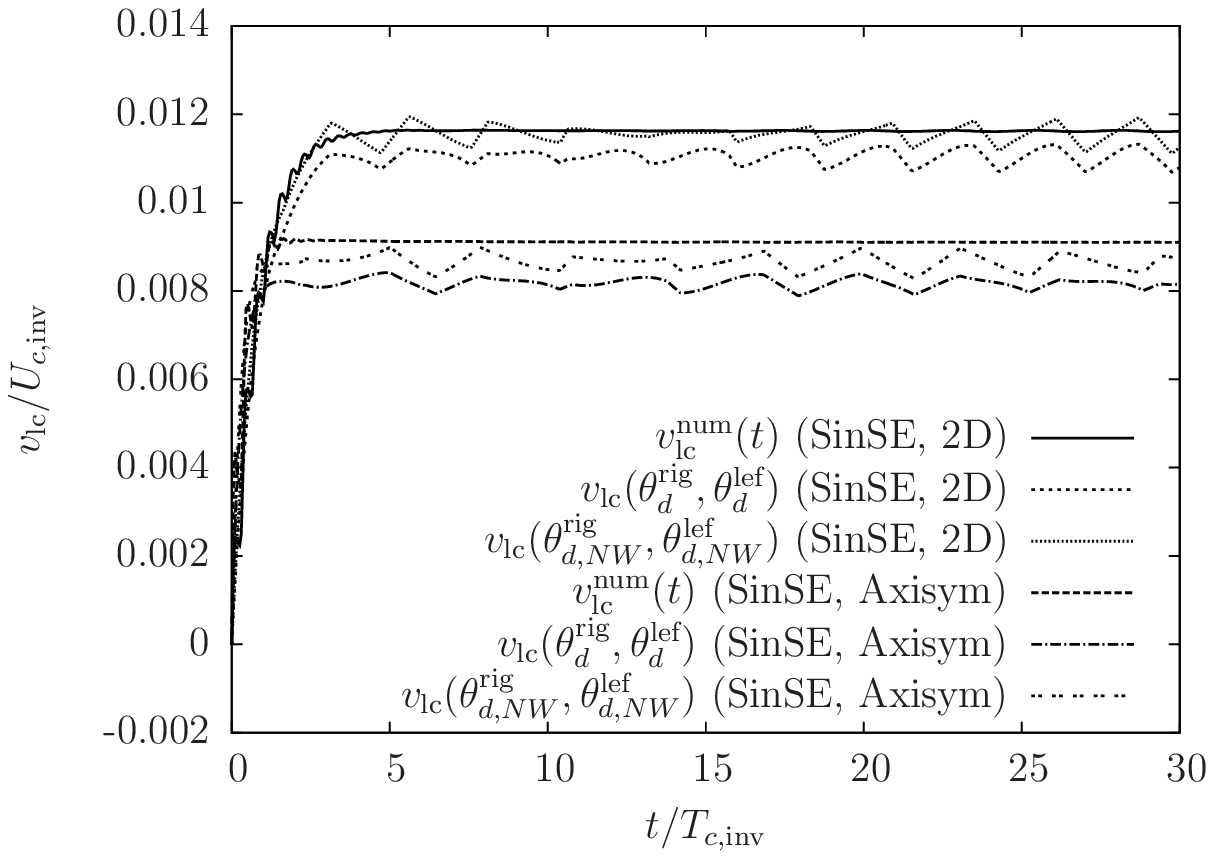}
(d)  \includegraphics[scale = 0.58]{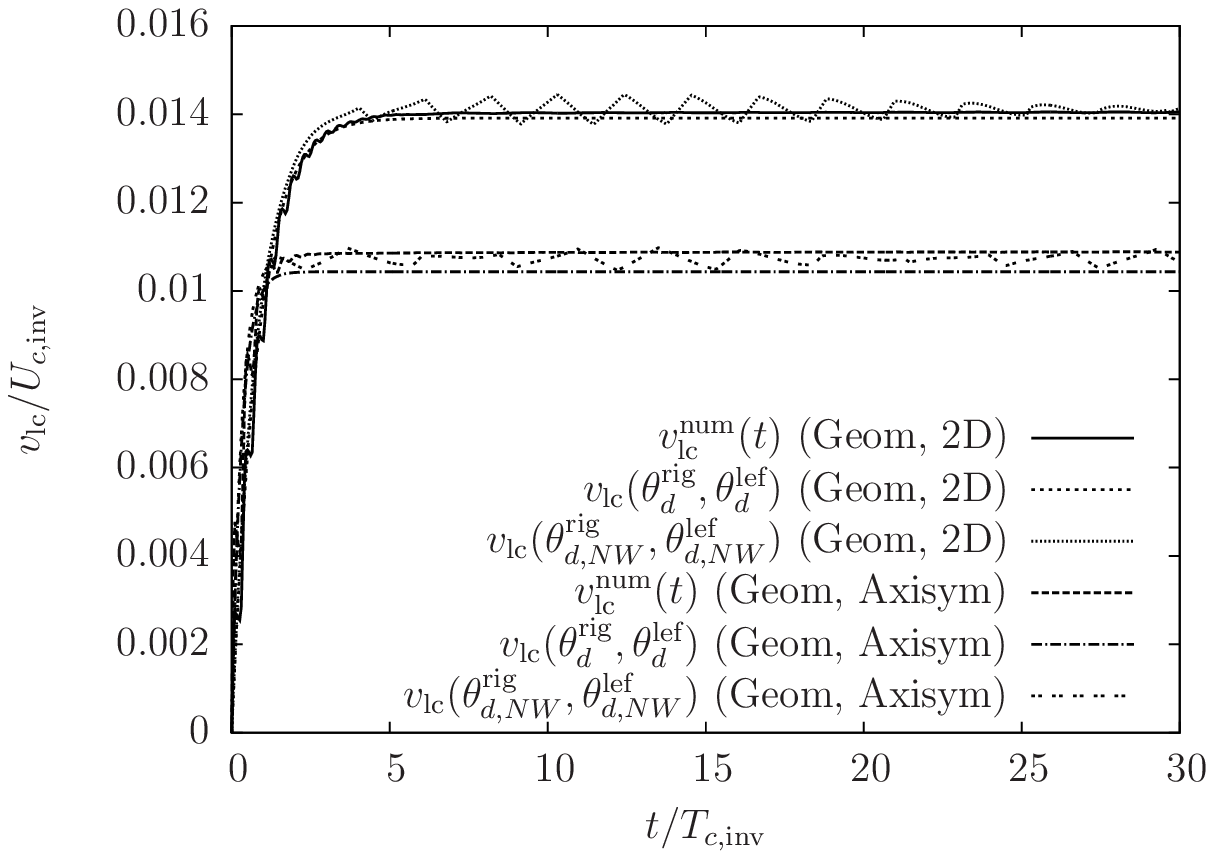}
(e)  \includegraphics[scale = 0.58]{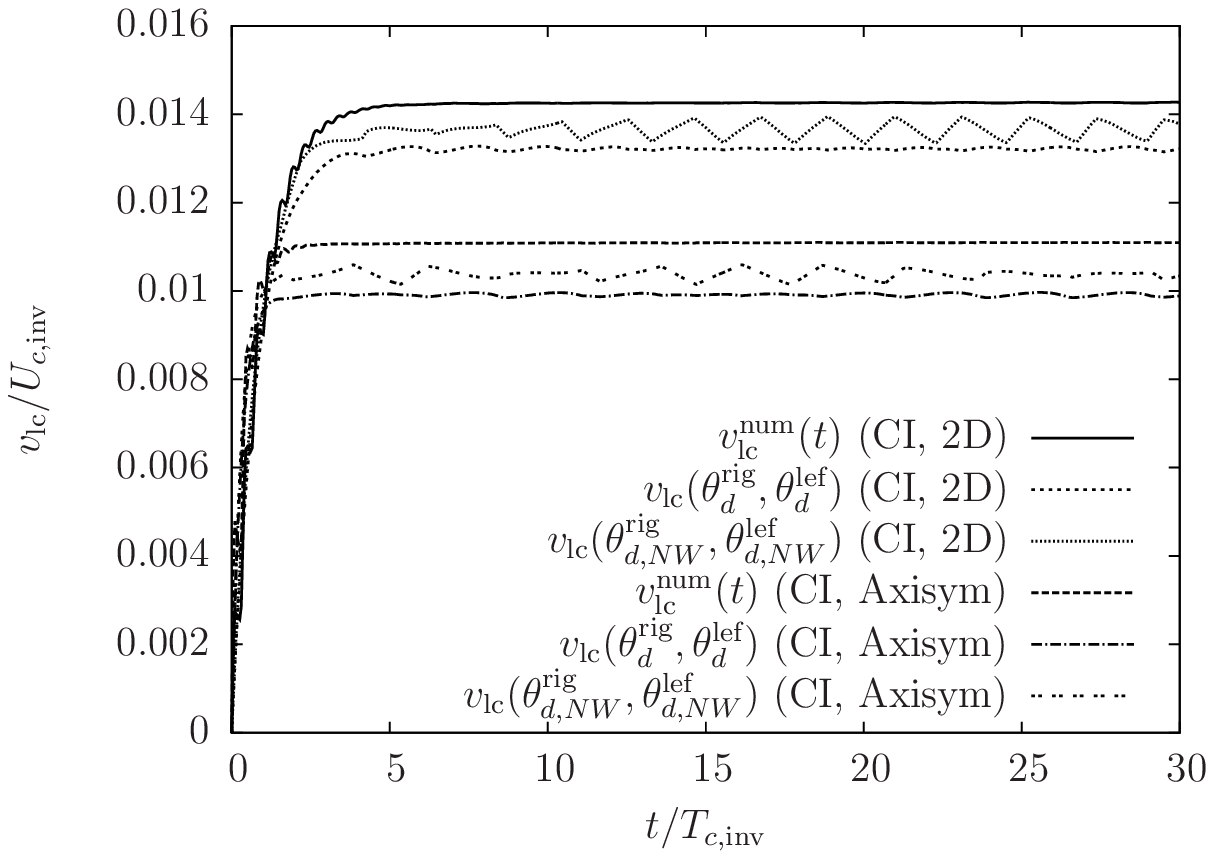}
  \caption{Comparison of the evolutions of the centroid velocity of the liquid column $v_{\textrm{lc}}$
  driven by a stepwise WG 
  under 2-D and cylindrical geometries 
  obtained by using the five different WBCs
  with those predicted by Eqs. (\ref{eq:vlc-theo}) and (\ref{eq:vlc-theo-axisym}) 
  for 2-D and axisymmetric problems, respectively:
  (a) LinSE; (b) CubSE; (c) SinSE; (d) Geom; (e) CI.
The common parameters are $L_{x} = 20$, $L_{y}= 0.5$, $Re = 100$ ($Oh = 0.1$), 
$\theta_{w}^{\textrm{rig}} = 47^{\circ}$, $\theta_{w}^{\textrm{lef}} = 59^{\circ}$,
$Cn=0.125$, $Pe = 5000$ ($S = 0.014$).}
  \label{fig:cmp-vlc-wg-rec}
\end{figure}

For this problem, all the five WBCs were tried (without wall energy relaxation, i.e., $r_{wr} = 1$).
The common parameters are 
$Re = 100$ ($Oh = 0.1$), $r_{\nu} = 1$ (i.e., $ \nu_{A} = \nu_{B}$),
$\theta_{w}^{\textrm{rig}} = 47^{\circ}$, $\theta_{w}^{\textrm{lef}} = 59^{\circ}$,
$Cn=0.125$, $Pe = 5000$ ($S = 0.014$), $N_{L}=32$, $N_{t}=320$.
Figure \ref{fig:cmp-vlc-wg-rec} shows 
the evolutions of the centroid velocity of the liquid column $v_{\textrm{lc}}$  
obtained by using the five different WBCs 
for $0 \leq t \leq 30 T_{c, \textrm{inv}}$.
Note that for this problem $U_{c, \textrm{inv}}$ and $T_{c, \textrm{inv}}$
were derived as in Eq. (\ref{eq:Uc-Tc-inv}), 
with the drop radius $R$ replaced by the channel height / tube diameter $H$. 
In Fig. \ref{fig:cmp-vlc-wg-rec}
the velocity and time are scaled by $U_{c, \textrm{inv}}$ and $T_{c, \textrm{inv}}$ respectively.
In actual simulations, the dynamic contact angles on the upper wall, 
$\theta_{d}^{\textrm{rig}}$ and $\theta_{d}^{\textrm{lef}}$,
were monitored and found to vary with time when LinSE, SinSE, CubSE and CI 
were used though the variations were within certain small ranges.
When Geom was used, these angles remained to be always exactly the same as the given ones, i.e.,
$\theta_{d}^{\textrm{rig}}=\theta_{w}^{\textrm{rig}}$ and $\theta_{d}^{\textrm{lef}}=\theta_{w}^{\textrm{lef}}$.
Also shown in Fig. \ref{fig:cmp-vlc-wg-rec}
are the theoretical predictions of $v_{\textrm{lc}}$ given by 
Eqs. (\ref{eq:vlc-theo}) and (\ref{eq:vlc-theo-axisym}) 
for 2-D and axisymmetric cases, respectively.
Note that in Fig. \ref{fig:cmp-vlc-wg-rec}
the dynamic contact angles ($\theta_{d}^{\textrm{rig}}$ and $\theta_{d}^{\textrm{lef}}$)
actually vary with time (except for Geom).
However, because the variations are small 
the plots may be still useful 
(as they provide rough estimates on the expected $v_{\textrm{lc}}$
based on the contact angles).
Besides, the dynamic contact angles \textit{near} the upper wall,
$\theta_{d, NW}^{\textrm{rig}}$ and $\theta_{d, NW}^{\textrm{lef}}$,
obtained one grid away from the wall 
(at $y = 0.5 H - h$) near the interfaces,
were also recorded and found to be time-dependent for all of the five WBCs.
The different velocities $v_{\textrm{lc}}$ calculated 
from $\theta_{d, NW}^{\textrm{rig}} (t)$ and $\theta_{d, NW}^{\textrm{lef}} (t)$
are also plotted in Fig. \ref{fig:cmp-vlc-wg-rec}.
Table \ref{tab:cmp-vlc-deviation-wg-rec} shows the deviations in $v_{\textrm{lc}}$
at the end of simulation ($t_{e} = 30 T_{c, \textrm{inv}}$),
when compared with the respective theoretical values
predicted by the two equations,
Eqs. (\ref{eq:vlc-theo}) and (\ref{eq:vlc-theo-axisym}), 
using the two sets of dynamic contact angles
(on the wall, $\theta_{d}^{\textrm{rig}}$ and $\theta_{d}^{\textrm{lef}}$,
and near the wall, $\theta_{d, NW}^{\textrm{rig}}$ and $\theta_{d, NW}^{\textrm{lef}}$)
for 2-D and axisymmetric problems, respectively.
The deviations were calculated as
$(v_{\textrm{lc}}^{\textrm{num}} - v_{\textrm{lc}} (\theta_{d}^{\textrm{rig}}, \theta_{d}^{\textrm{lef}}) )/
v_{\textrm{lc}} (\theta_{d}^{\textrm{rig}}, \theta_{d}^{\textrm{lef}}) \times 100 \%$
and $(v_{\textrm{lc}}^{\textrm{num}} - v_{\textrm{lc}} (\theta_{d, NW}^{\textrm{rig}}, \theta_{d, NW}^{\textrm{lef}}) )/
v_{\textrm{lc}} (\theta_{d, NW}^{\textrm{rig}}, \theta_{d, NW}^{\textrm{lef}}) \times 100 \%$.
From Fig. \ref{fig:cmp-vlc-wg-rec}
and Table \ref{tab:cmp-vlc-deviation-wg-rec}, it is seen that 
the results by all WBCs (except for LinSE)
agree roughly (in many cases, reasonably well)
with the respective theoretical predictions using the dynamic contact angles.
When the dynamic contact angles \textit{near the wall} are used,
the agreement seems to improve for all cases considered.  
The deviations by LinSE are the largest among all.
This could be due to that the dynamic contact angles were not well captured
because the parallel contours of the order parameter in the interfacial region
were not well preserved when LinSE was used (as will be shown later).
The deviations by Geom are the smallest (less than $5 \%$ for both 2-D
and axisymmetric cases using either set of dynamic contact angles).
The other three WBCs (CubSE, SinSE and CI) 
had comparable performances in terms of the deviation in $v_{\textrm{lc}}$ 
(the maximum deviations are around $10 \%$).
It can also be found from Fig. \ref{fig:cmp-vlc-wg-rec}
that in all cases the time for the liquid column to reach nearly steady state
agrees well with the theoretical predictions except for LinSE.

\begin{table}
 \centering
\begin{tabular}{|c|c|c|c|c|c|}\hline
WBC & LinSE & CubSE & SinSE & Geom & CI \\ \hline
Deviation in $v_{\textrm{lc}}$ (2D, using $\theta_{d}$) & 57.8 \% & 5.9 \% & 7.6 \% & 1.0 \% & 8.0 \% \\\hline
Deviation in $v_{\textrm{lc}}$ (2D, using $\theta_{d, NW}$) & 31.1 \% & 3.4 \% & 3.2 \% & -0.5 \% & 3.6 \% \\\hline
Deviation in $v_{\textrm{lc}}$ (Axisym, using $\theta_{d}$) & 49.9 \% & 12.2 \% & 11.6 \% & 4.3 \% & 12.2 \% \\\hline
Deviation in $v_{\textrm{lc}}$ (Axisym, using $\theta_{d, NW}$) & 35.9 \% & 4.8 \% & 4.1 \% & 2.2 \% & 7.3 \% \\\hline
\end{tabular}
 \caption{Deviations of the centroid velocity of the liquid column 
 $v_{\textrm{lc}}$ at $t_{e} = 30 \ (T_{c, \textrm{inv}})$
  driven by a stepwise WG 
  under 2-D and cylindrical geometries 
  obtained by using five different WBCs
  when compared with those predicted by Eqs. (\ref{eq:vlc-theo}) and (\ref{eq:vlc-theo-axisym}) 
  using ($\theta_{d}^{\textrm{rig}} (t_{e} )$, $\theta_{d}^{\textrm{lef}} (t_{e} )$)
  and ($\theta_{d, NW}^{\textrm{rig}} (t_{e} )$, $\theta_{d, NW}^{\textrm{lef}} (t_{e} )$)
  respectively.}
  \label{tab:cmp-vlc-deviation-wg-rec}
\end{table}

\begin{figure}[htp]
  \centering
(a)  \includegraphics[scale = 0.58]{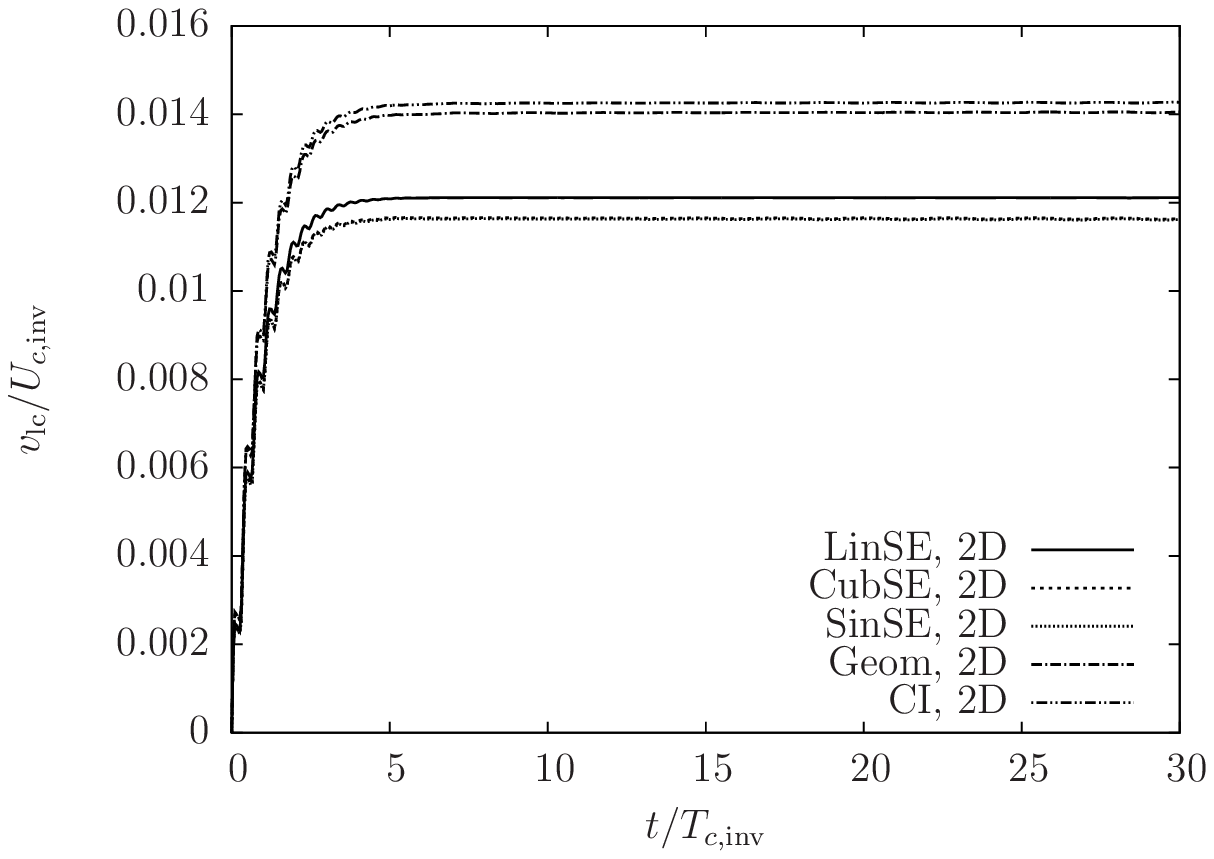}
(b)  \includegraphics[scale = 0.58]{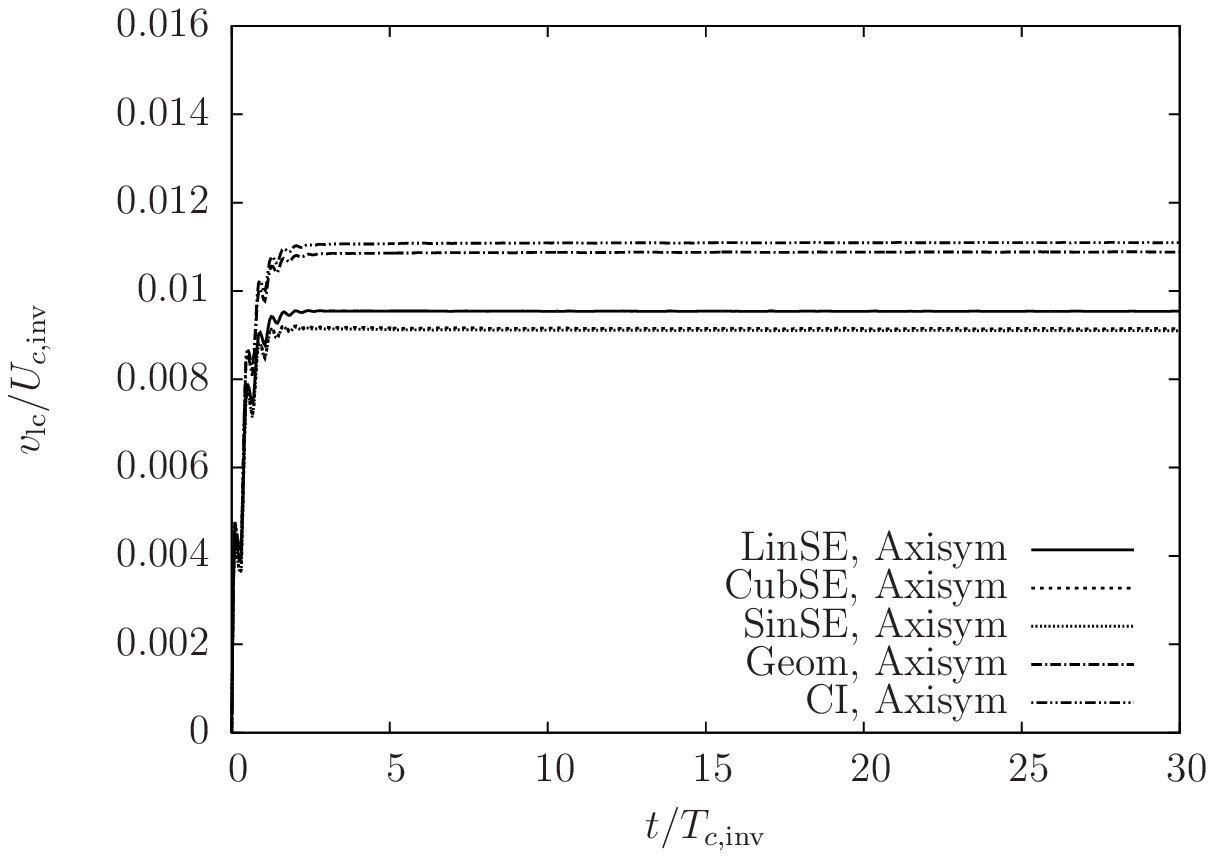}
  \caption{Comparison of the evolutions of the liquid column velocity $v_{\textrm{lc}}$
  driven by a stepwise WG 
  under (a) 2-D and (b) cylindrical geometries 
  obtained by using five different WBCs.
The common parameters are $L_{x} = 20$, $L_{y}= 0.5$, $Re = 100$ ($Oh = 0.1$), 
$\theta_{w}^{\textrm{rig}} = 47^{\circ}$, $\theta_{w}^{\textrm{lef}} = 59^{\circ}$,
$Cn=0.125$, $Pe = 5000$ ($S = 0.014$).}
  \label{fig:cmp-vlc-wg-rec-wbc-2d-axisym}
\end{figure}

It would be helpful to make some comparisons between different WBCs as well.
Figure \ref{fig:cmp-vlc-wg-rec-wbc-2d-axisym}
compares the evolutions of $v_{\textrm{lc}}$
by using different WBCs under 2-D and axisymmetric geometries.
Table \ref{tab:cmp-vlc-wbc-wg-rec} gives the details of $v_{\textrm{lc}}$
at the end of simulation by using all five WBCs.
It is obvious that the steady velocity in 2-D is in general greater than
that in axisymmetric geometry for each of the WBCs.
Under both geometries, 
CubSE and SinSE almost give
identical evolutions of $v_{\textrm{lc}}$
and the steady velocities by them are the smallest among all.
The results by Geom and CI are very close to each other as well
and their predictions of the steady velocities are larger than those by the others.
This may be due to the fact that both Geom and CI aim to enforce the local contact angles 
on the wall to be the static ones.
Finally, the predictions by LinSE lie in between those by the other two groups.

\begin{table}
 \centering
\begin{tabular}{|c|c|c|c|c|c|}\hline
WBC & LinSE & CubSE & SinSE & Geom & CI \\ \hline
$v_{\textrm{lc}} / U_{c, \textrm{inv}}$ (2D) & 0.0121 & 0.0116 & 0.0116 & 0.0141 & 0.0143 \\\hline
$v_{\textrm{lc}} / U_{c, \textrm{inv}}$ (Axisym) &  0.0095 & 0.0091 & 0.0091 & 0.0109 & 0.0111 \\\hline
\end{tabular}
 \caption{Comparison of the (steady) velocity of the liquid column 
 $v_{\textrm{lc}}$ (measued in $U_{c, \textrm{inv}}$) at $t_{e} = 30 \ (T_{c, \textrm{inv}})$
  driven by a stepwise WG 
  under 2-D and cylindrical geometries 
  obtained by using five different WBCs.}
  \label{tab:cmp-vlc-wbc-wg-rec}
\end{table}

Based on the above comparisons between the simulated results and theoretical predictions
as well as between those by different WBCs, it may be concluded that 
(1) all WBCs, except LinSE, can give fairly consistent results of the velocity and the dynamic contact angles
for the motion of a liquid column driven by a stepwise WG;
(2) the use of different WBCs can lead to noticeable differences in the results of the velocity;
(3) among the five WBCs considered, CubSE and SinSE give very close results
whereas Geom and CI belong to another group;  
(4) although the consistency in the results by LinSE is not quite satisfactory,
it could be still used to predict the velocity.

\subsection{Study of a Drop Dewtting from a Lyophobic Surface}\label{ssec:dewetting}

The third dynamic problem is the dewetting of a drop from a lyophobic surface under cylindrical geometry.
Drop dewetting may appear during the dewetting of a thin liquid film on
a nonwetting surface~\cite{rmp09-wetting-spreading, prl04dewetting};
it may also be encountered when a drop sits on a surface 
with dynamically changing wettability,
for instance, controlled by electric field in microfluidic devices
~\cite{apl06ew-mixing, jast-hybrid-lbm-fvm}.
While drop spreading has been investigated heavily
~\cite{DIMSpreading07, pre07-geom-wbc, jfm07-spreading-di-ls,
pof09-dyn-wetting, pof11-wall-energy-relax},
studies on drop dewetting are relatively scarce
(Huang et al. did some studies with LBM on drop dewetting~\cite{ijnmf09-pflbm-mobility},
but their study focused mainly on two-dimensional (2-D) problems and only used LinSE).
Therefore, the present study has significance because it
provides some insights not only on various WBCs in PF simulations
but also on the important problem of drop dewetting.

As mentioned before, the problem setup in the study of drop dewetting is the same as that 
in the study of static contact angle in Section \ref{ssec:staticca}.
Based on the recorded drop \textit{radius} on the left wall $R_{y} (t)$,
one can calculate the contact line velocity $V_{\textrm{cl}}$ by using backward differentiation.
For instance, $V_{\textrm{cl}}^{t}$ at time $t$ is found from,
\begin{equation}
  \label{eq:CLVel}
 	V_{\textrm{cl}}^{t} = \frac{1}{\Delta t} (R_{y}^{t} - R_{y}^{t-\Delta t})  ,
\end{equation}
where $\Delta t$ is the change in time and may take $k \delta_{t}$
with $k$ being a positive integer (note $V_{\textrm{cl}}^{t}$
is more smoothed out at larger $\Delta t$).
Because at $t=0$ the initial contact line velocity $V_{\textrm{cl}}^{0}$ can not be calculated by the above formula,
we assume $V_{\textrm{cl}}^{0} \approx V_{\textrm{cl}}^{1}$
(in other words, the forward differentiation is employed to calculate $V_{\textrm{cl}}^{0}$.)
We also look into the local dynamic contact angle.
Since the interfacial region covers a few grid points in PF simulations, 
it may occur that the local contact angle $\theta_{d, l}$ obtained from Eq. (\ref{eq:LocalCA})
varies across these grid points. 
The maximum and minimum values of $\theta_{d, l}$
across the interfacial region, 
$\theta_{d, l}^{\textrm{max}}$ and $\theta_{d, l}^{\textrm{min}}$, 
are recorded at each time step for the drop dewetting problem.
For the flow field, we monitor
the maximum velocity magnitude over the domain
at time $t$ defined as,
\begin{equation}
  \label{eq:vm-def}
  \sqrt{u^2+v^2}\vert _{\textrm{max}} (t)
  = \max_{i,j} \sqrt{(u_{i,j}(t))^2+(v_{i,j}(t))^2}   .
\end{equation}

\subsubsection{Effects of the Cahn number and Peclect number}

As noted in Subsection \ref{ssec:char-quant-diml-num-setup},
the Cahn number $Cn$ and the Peclect number $Pe$
(or the parameter $S$) are two additional parameters in PF
simulations.
In the study of previous problems, we simply used some suitable values for these numbers.
In this section, we carry out some studies on the effects of these two parameters
for the problem of drop dewetting.
For conciseness, we only consider one of the WBCs
for the study of the effects of $Cn$ and $Pe$ ($S$).

\begin{figure}[htp]
  \centering
	(a) \includegraphics[scale = 0.60]{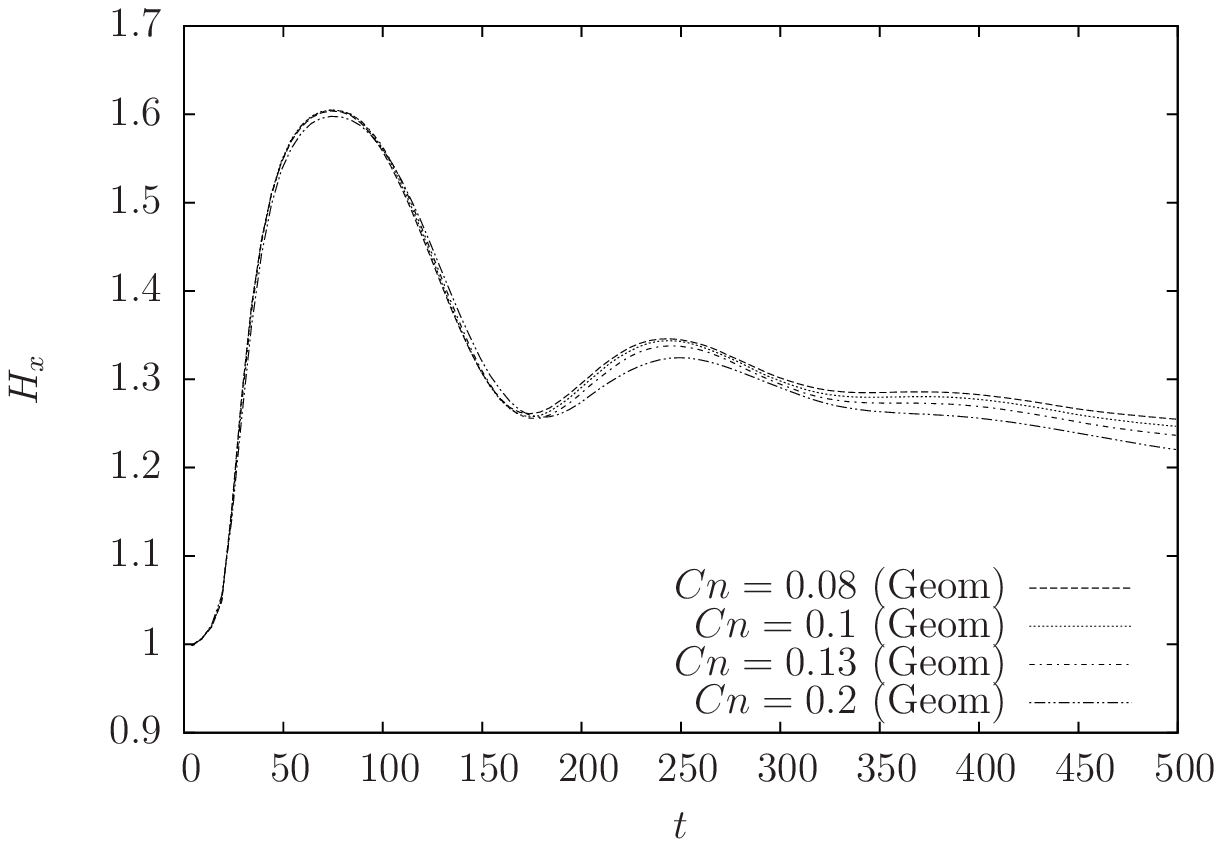}	
	(b) \includegraphics[scale = 0.60]{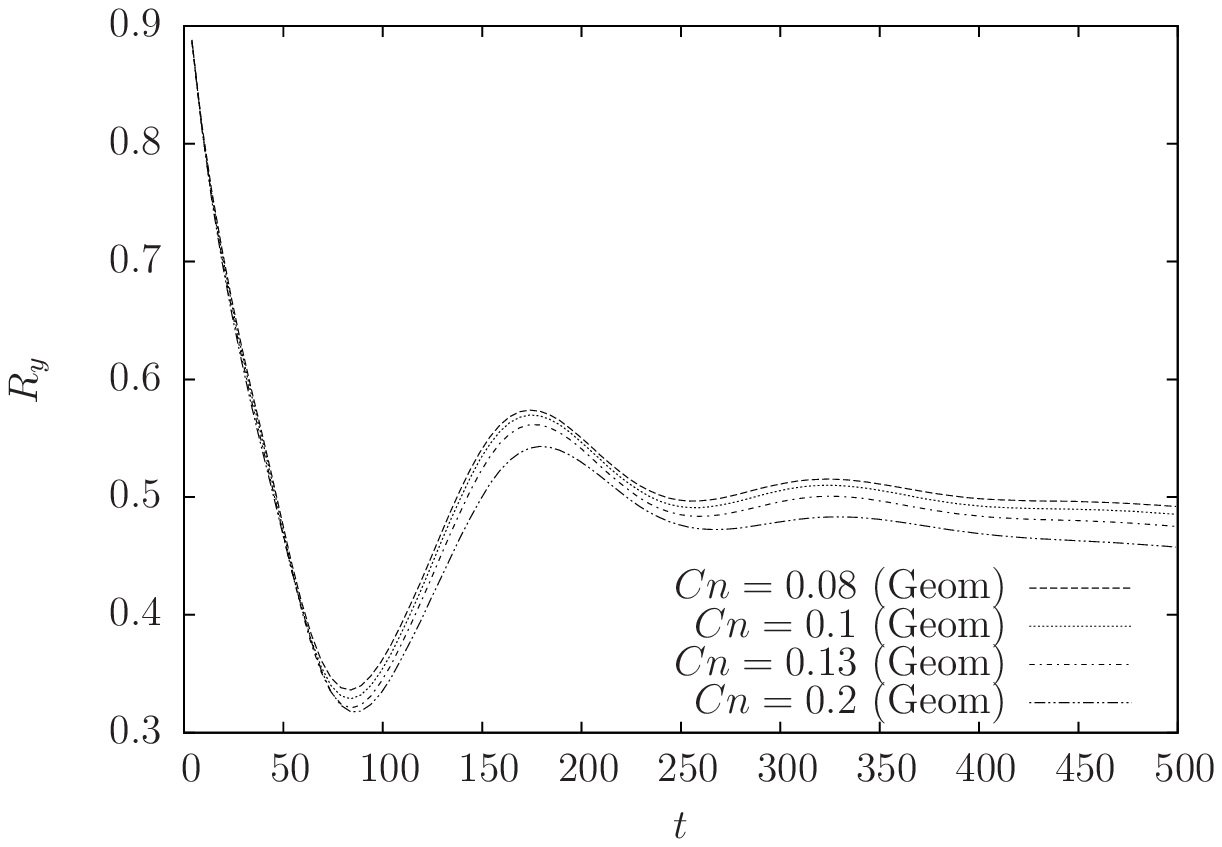}
	(c) \includegraphics[scale = 0.58]{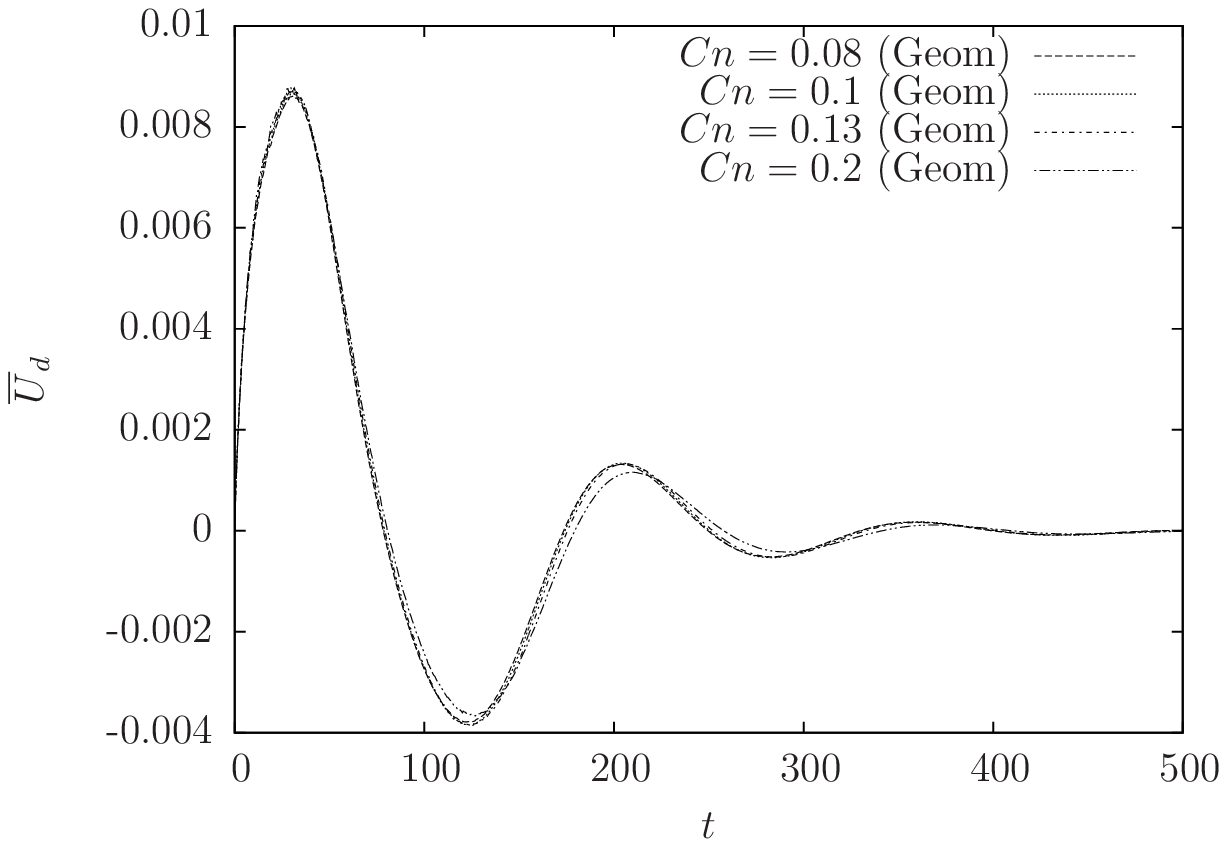}
	(d) \includegraphics[scale = 0.58]{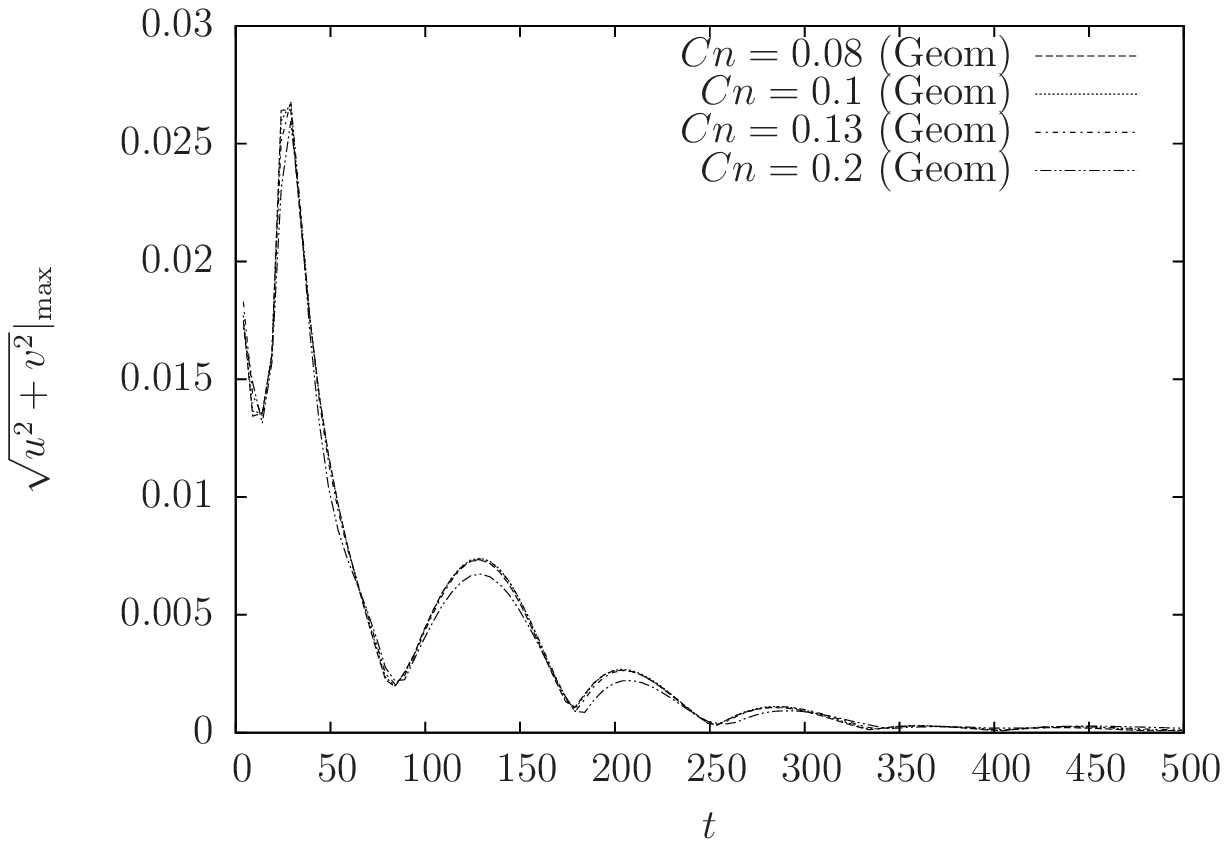}	
  \caption{Comparison of 
  (a) the drop \emph{height} on the $x-$axis $H_{x}$
 (b) the drop \textit{radius} on the left wall $R_{y}$  
  (c) the average drop velocity $\overline{U}_{d}$ in the $x-$direction 
  (d) the maximum velocity magnitude $\sqrt{u^2+v^2}\vert _{\textrm{max}}$ 
  under different Cahn numbers for $\theta_{w} = 135^{\circ}$, $\theta_{i} = 90^{\circ}$,
  $Re=1000$ ($Oh = 0.032$), $Ca=1$ and $Pe=5000$ ($S = 0.014$) by using Geom.}
  \label{fig:cn-cmp-ud-vm-hx-ry-Re1k-ca135-Pe5k}
\end{figure}

For macroscopic continuum simulations, $Cn$ should be ideally zero.
But in actual simulations, this is impossible and $Cn$ must take some finitely small value.
Fortunately, for a small enough $Cn$ the simulation results
can be reasonably close to the \textit{sharp-interface limit}~\cite{jfm10-sil-che-cl}.
Here we pick one case with these physical parameters: 
the wettability of the left wall is specified by a (static) contact angle $\theta_{w} = 135^{\circ}$, 
the initial configuration corresponds to an initial contact angle $\theta_{i} = 90^{\circ}$, 
the Reynolds number (Ohnesorge number) and capillary number 
are $Re=1000$ ($Oh = 0.032$) and $Ca=1$.
We fix the Peclet number at $Pe=5000$ ($S = 0.014$),
and perform a series of simulations using different Cahn numbers
to investigate its effects. 
The WBC uses the geometric formulation (Geom).
Figure \ref{fig:cn-cmp-ud-vm-hx-ry-Re1k-ca135-Pe5k} compares
several average, extreme or local quantities, including 
the drop \emph{height} on the $x-$axis $H_{x}$,
the drop \textit{radius} on the left wall $R_{y}$,
the average drop velocity $\overline{U}_{d}$ in the $x-$direction, 
and the maximum velocity magnitude $\sqrt{u^2+v^2}\vert _{\textrm{max}}$
under four Cahn numbers, $Cn=0.2, \ 0.13, \ 0.1, \ 0.08$
(corresponding to $N_{L} = 20, \ 30, \ 40, \ 50$ while fixing $\tilde{W} = 4$),
for the selected case.
It is first helpful to analyze the basic dewetting process.
Because the initial configuration corresponds to $\theta_{i} = 90^{\circ}$,
which is less than $\theta_{w} = 135^{\circ}$,
the surface stresses along the wall are not balanced.
This initial unbalance of force drives the drop towards the equilibrium configuration,
making $R_{y}$ decrease and $H_{x}$ increase during the early stage,
as found in Fig. \ref{fig:cn-cmp-ud-vm-hx-ry-Re1k-ca135-Pe5k}a\&b.
At the same time, the drop gradually gains certain momentum in the $x-$direction
(see Fig. \ref{fig:cn-cmp-ud-vm-hx-ry-Re1k-ca135-Pe5k}c),
and the maximum velocity magnitude increases very quickly
(see Fig. \ref{fig:cn-cmp-ud-vm-hx-ry-Re1k-ca135-Pe5k}d).
After some time, $H_{x}$ reaches a peak and starts to decrease,
and subsequently experiences some oscillations with the amplitude slowly decaying.
Decayed oscillations are also seen in the evolution of $R_{y}$, $\overline{U}_{d}$
and $\sqrt{u^2+v^2}\vert _{\textrm{max}}$.
From Fig. \ref{fig:cn-cmp-ud-vm-hx-ry-Re1k-ca135-Pe5k},
it is found that as $Cn$ decreases the local quantities ($H_{x}$ and $R_{y}$)
show relatively large changes (not during the early stage but after certain time,
e.g., for $H_{x}$, $t > 150$, and for $R_{y}$, $t > 75$).
We note that in the late stages both $H_{x}$ and $R_{y}$ decrease,
which means the drop shrinks.
This is related to the intrinsic property of the PF model
and closely tied to $Cn$~\cite{jcp07pf-drop-shrink, DIMSpreading07}.
Thus, the differences in $H_{x}$ and $R_{y}$ under different $Cn$ are more related
to this shrinkage rather the dewetting process.
What is more, as $Cn$ decreases the differences also decrease. 
Such drop shrinkage should be acceptable as long as 
it is controlled to a certain extent during the course of simulation.
Unlike the local quantities, the average and extreme quantities 
($\overline{U}_{d}$ and $\sqrt{u^2+v^2}\vert _{\textrm{max}}$)
change only a little with the variation of $Cn$;
in fact, noticeable differences are found only between $Cn=0.2$ and $Cn=0.13$
for these two quantities.
To have a good balance between the computational cost
and the accuracy of the results, we use $Cn=0.1$ for most of the simulations in the study of drop dewetting.

\begin{figure}[htp]
  \centering
  \includegraphics[scale = 0.8]{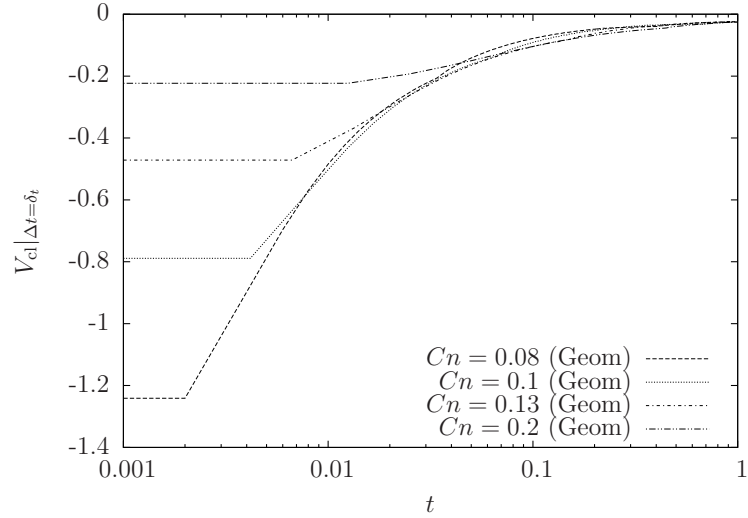}
  \caption{Comparison of the evolutions of the contact line velocity 
  $V_{\textrm{cl}}$ obtained at $\Delta t = \delta_{t}$
  in the early stage $0 < t \leq 1$ (with $t$ plotted in log scale)
  under different Cahn numbers
  for $\theta_{w} = 135^{\circ}$, $\theta_{i} = 90^{\circ}$,
  $Re=1000$ ($Oh = 0.032$), $Ca=1$ and $Pe=5000$ ($S = 0.014$) by using Geom.}
    \label{fig:cn-cmp-vcl-Re1k-ca135-Pe5k}
\end{figure}

At the beginning, the unbalance of the surface forces along the wall 
is the strongest. As found in Fig. \ref{fig:cn-cmp-ud-vm-hx-ry-Re1k-ca135-Pe5k}b,
initially $R_{y}$ decreases very sharply.
Thus, we are especially interested in the contact line velocity during the early stage.
Figure \ref{fig:cn-cmp-vcl-Re1k-ca135-Pe5k} 
shows the contact line velocity $V_\textrm{cl}$ obtained with $\Delta t = \delta_{t}$
in the early stage $0 < t \leq 1$ 
under different Cahn numbers: $Cn=0.2, \ 0.13, \ 0.1, \ 0.08$.
From Fig. \ref{fig:cn-cmp-vcl-Re1k-ca135-Pe5k}, one finds that
the magnitude of  $V_{\textrm{cl}}$ at the beginning increases
as $Cn$ decreases, and such a change with $Cn$ is in fact quite noticeable
(for example, from about $0.8$ at $Cn=0.1$ to about $1.2$ at $Cn=0.08$).
Although large differences are observed in $V_{\textrm{cl}}$ 
at the beginning, 
such differences decay very fast and become almost negligible at $t=1$.
Besides, they seem to have no significant effect on subsequent drop motion,
as seen in Fig. \ref{fig:cn-cmp-ud-vm-hx-ry-Re1k-ca135-Pe5k}.
Thus, unless it is necessary to resolve the details of drop dewetting at the start,
it should be acceptable to use $Cn=0.1$.

\begin{figure}[htp]
  \centering
(a)  \includegraphics[scale = 0.58]{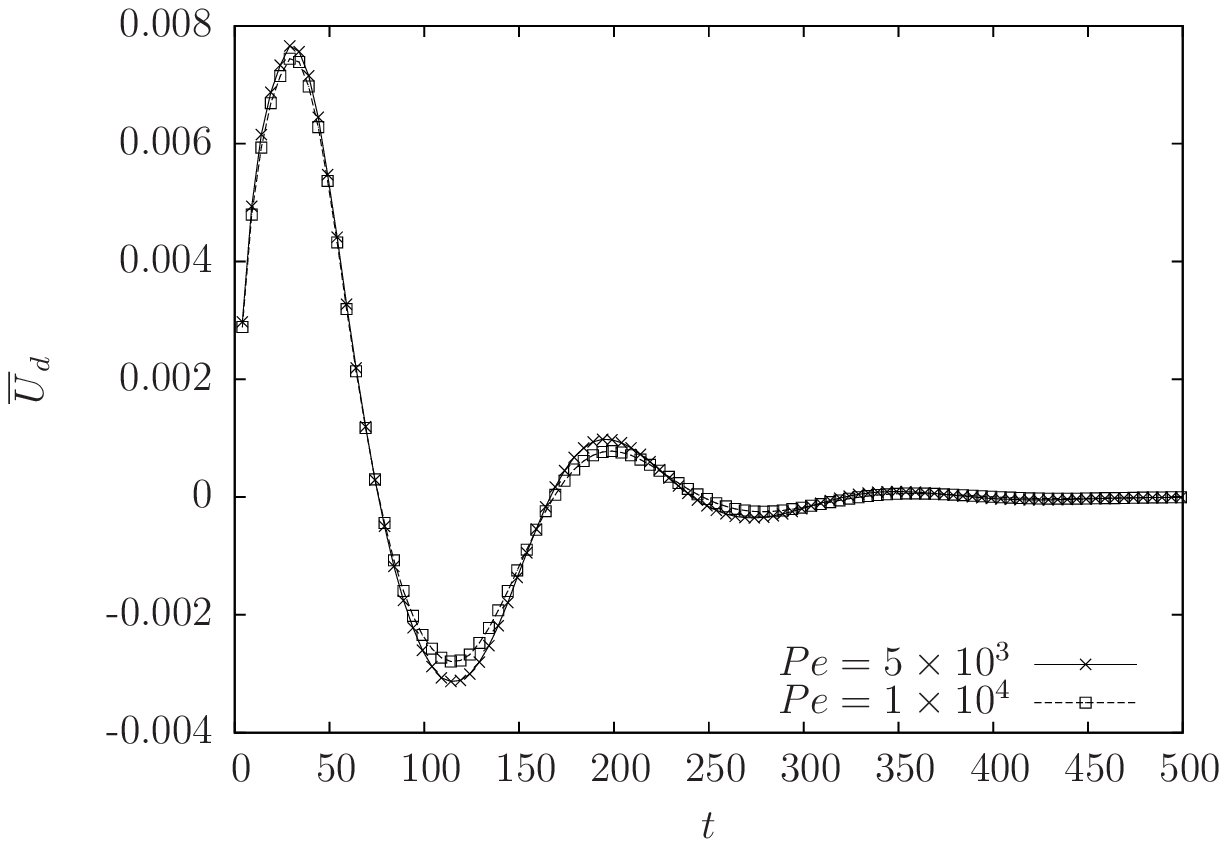}
(b)  \includegraphics[scale = 0.58]{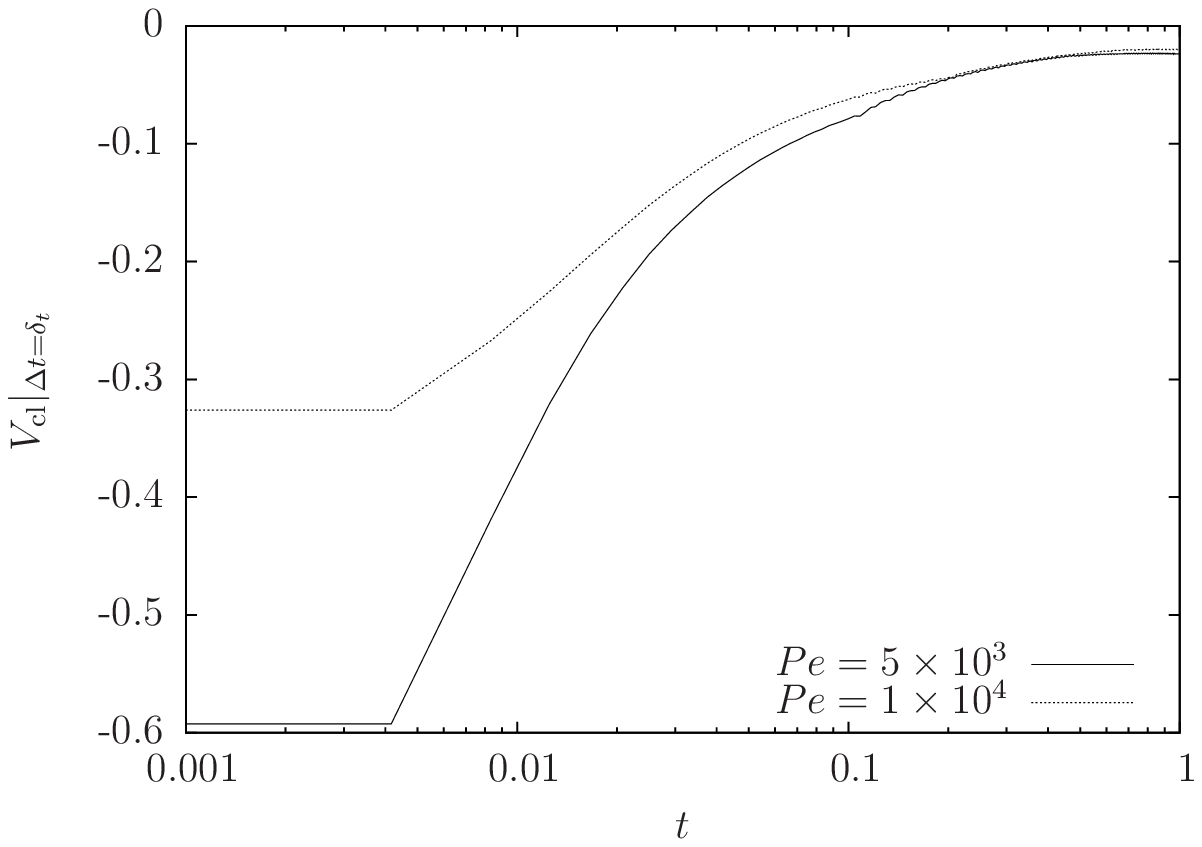}
  \caption{Comparison of the evolutions of
    (a) the average drop velocity $\overline{U}_{d}$ in the $x-$direction 
    (b) the contact line velocity $V_{\textrm{cl}}$ obtained at $\Delta t = \delta_{t}$
    (in the early stage $0 < t \leq 1$ with $t$ plotted in log scale)
  under different Peclet numbers for $\theta_{w} = 135^{\circ}$, $\theta_{i} = 90^{\circ}$, 
  $Re=1000$ ($Oh = 0.032$), $Ca=1$,
  $Cn=0.1$ by using CubSE.}
  \label{fig:pe-cmp-ud-vcl-Re1k-ca135-cn0d1}
\end{figure}

Next, the effects of $Pe$ (or the parameter $S$) are briefly investigated 
while $Cn$ is fixed at $0.1$.
It is already known that $Pe$ controls the diffusion in the CHE.
In our study, it was found that when $Pe$ is too small (i.e., the diffusion is too strong),
the drop shrinkage rate becomes too large to be acceptable.
At the same time, $Pe$ must not be too large (to keep sufficient diffusion)
so that the profile of $\phi$ in the interfacial region
can be well maintained~\cite{jacqmin99jcp}.
Besides, it has been found that the diffusion length scale at the contact line is related to $Pe$ ($S$)
~\cite{jfm10-sil-che-cl}.
Here we do not intend to dig into this issue
and only consider two values of $Pe$ ($S$) (which we deem to be in the suitable range): 
$Pe=5000$ ($S = 0.014$) and $Pe=10000$ ($S = 0.01$).
For the study of $Pe$'s effects, the WBC employed is CubSE.
Figure \ref{fig:pe-cmp-ud-vcl-Re1k-ca135-cn0d1} 
compares the evolution of 
(a) the average drop velocity $\overline{U}_{d}$ in the $x-$direction 
and (b) the contact line velocity $V_{\textrm{cl}}$ (in the early stage $0 < t \leq 1$)
under the two Peclet numbers.
It is observed from Fig. \ref{fig:pe-cmp-ud-vcl-Re1k-ca135-cn0d1}a 
that the results on $\overline{U}_{d}$ obtained with $Pe=5000$ and $Pe=10000$
are overall very close to each other.
The evolution of $\overline{U}_{d}$ with $Pe=5000$ (larger diffusion) shows
slightly larger amplitude of oscillation (i.e., larger maximum and smaller minimum).
From Fig. \ref{fig:pe-cmp-ud-vcl-Re1k-ca135-cn0d1}b one finds that,
similar to what is seen in Fig. \ref{fig:cn-cmp-vcl-Re1k-ca135-Pe5k},
$Pe$ (or $S$) affects the contact line velocity
at the beginning quite significantly:
$V_{\textrm{cl}}^{0}$ at $Pe=5000$ ($S = 0.014$) is almost twice of 
that at $Pe=10000$ ($S = 0.01$).
At the same time, the difference in $V_{\textrm{cl}}$ 
caused by the change of $Pe$ decays quickly with time,
and could nearly be ignored at $t=1$.
To reduce the dimension of the parameters
we use $Pe=5000$ ($S = 0.014$) below. 
However, it is noted that under certain circumstances
(not encountered here), $Pe$ could become a critical parameter to affect the final results 
(e.g., see~\cite{ijnmf09-pflbm-mobility}).
  
\subsubsection{Effects of the WBC}

In this part, we study how the use of different WBCs affects the outcome of simulation of drop dewetting.
The Cahn number is fixed at $Cn=0.1$
and the Peclet number is fixed at $Pe=5000$ ($S=0.014$).
The initial contact angle is $\theta_{i} = 90^{\circ}$, and 
the (static) contact angle of the left wall is $\theta_{w} = 135^{\circ}$.

\begin{figure}[htp]
  \centering
  (a) \includegraphics[scale = 0.6]{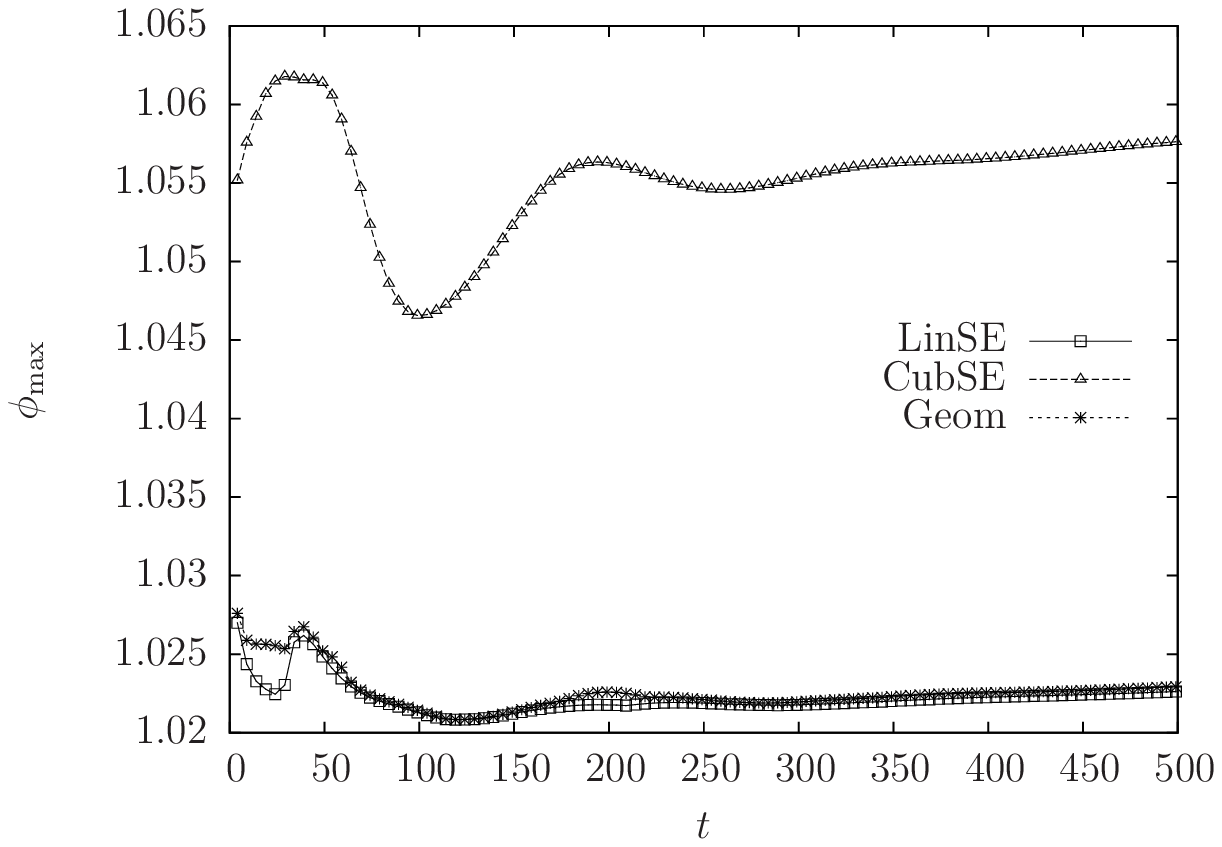}
  (b) \includegraphics[scale = 0.6]{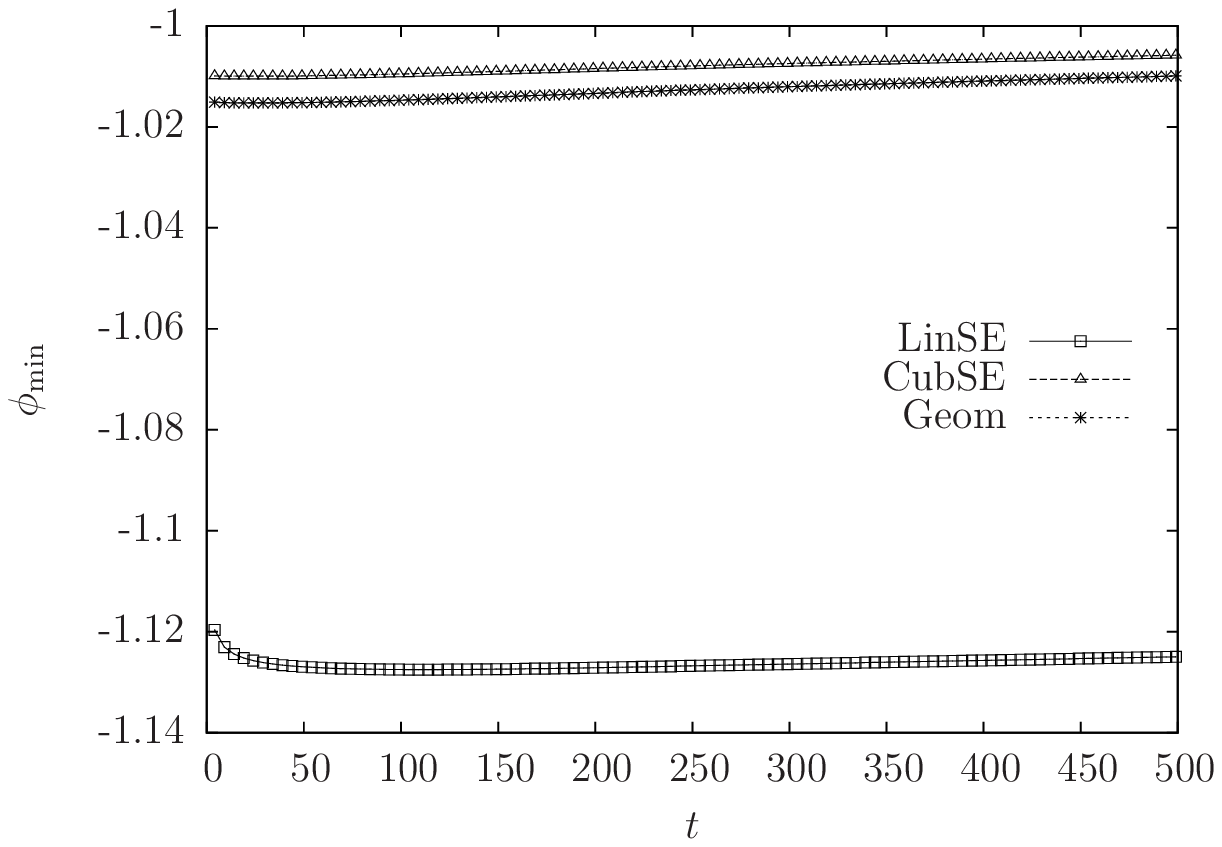}
  \caption{Comparison of the evolutions of $\phi_{\textrm{max}}$
  and $\phi_{\textrm{min}}$ by using different WBCs.
  The parameters are $\theta_{w} = 135^{\circ}$, $\theta_{i} = 90^{\circ}$, 
  $Re=1000$ ($Oh = 0.032$), $Ca=1$,
  $Cn=0.1$, $Pe=5000$ ($S = 0.014$), $N_{L} = 40$, $N_{t} = 240$.}
  \label{fig:cmp-phi-max-min-Re1k-ca135-cn0d1-Pe5k}
\end{figure}

We study a case at a relatively large Reynolds number, $Re=1000$ 
($Oh = 0.032$),
by using all five WBCs (LinSE, CubSE, SinSE, Geom and CI).
First, we found that the key observables of interest by using CubSE and SinSE
were very close to each other for this problem of drop dewetting;
besides, the results by using Geom and CI just had small differences.
Thus, we will focus only on three WBCs
(LinSE, CubSE and Geom).
As mentioned in Section \ref{sec:intro},
many researchers prefer to use CubSE rather than LinSE
because it avoids the appearance of the \textit{wall layer}
for $\theta_{w} \neq 90^{\circ}$.
We note that in general PF simulations $\phi$ may 
deviate from its equilibrium values $\pm 1$~\cite{jcp07pf-drop-shrink},
but the deviation is usually small under a small $Cn$ (e.g., $Cn=0.1$).
However, a wall layer could appear if LinSE is used and then $\phi$ on the wall
could take values that deviate much more from its equilibrium values ($\pm 1$)
when $\theta_{w}$ is far away from $90^{\circ}$.
For instance, on a lyophobic wall ($\theta_{w} > 90^{\circ}$),
the analytical prediction for $\phi$ is $\phi_{w} = - \sqrt{1 + \vert \tilde{\omega} \vert}$
if $\phi = -1$ in the bulk fluid~\cite{tipm02, mythesis09}.
For $\theta_{w} = 135^{\circ}$, it is found that $\phi_{w} \approx -1.21$.
As key indicators of this wall layer, the maximum and minimum values of $\phi$,
$\phi_{\textrm{max}}$ and $\phi_{\textrm{min}}$,
are monitored during the simulation.
Figure \ref{fig:cmp-phi-max-min-Re1k-ca135-cn0d1-Pe5k} compares
the evolutions of $\phi_{\textrm{max}}$ and $\phi_{\textrm{min}}$ by using different WBCs.
It is seen from Fig. \ref{fig:cmp-phi-max-min-Re1k-ca135-cn0d1-Pe5k}b
that $\phi_{\textrm{min}}$ by using LinSE deviates from $-1$ much more ($>10\%$)
than those by using CubSE and Geom ($<2\%$).
From Fig. \ref{fig:cmp-phi-max-min-Re1k-ca135-cn0d1-Pe5k}a
one finds that $\phi_{\textrm{max}}$ by using any of the three WBCs does not exceed $10\%$
and the one with CubSE deviates from $1$ relatively more (about $5\%$ to $6\%$).
A further examination of the contour plots (not shown here)
indicates that the large deviation in $\phi_{\textrm{min}}$ by using LinSE
indeed occurred inside the wall layer
whereas both CubSE and Geom were free from such wall layers.
The observed value $\phi_{\textrm{min}} \approx -1.12$ (for LinSE) is slightly larger
than theoretical one $-1.21$, but one should notice that
$\phi_{w}$ is to be taken exactly on the wall ($0.5 h$ away from the
outmost computational nodes)
and the normal gradient of $\phi$ is large for $\theta_{w} = 135^{\circ}$.
Some previous studies have found that
the appearance of the wall layer could make the simulations less accurate~\cite{cpc11wbc-lbm}.
We note that for binary fluids with non-equal densities and/or viscosities
larger deviations in $\phi$ from its equilibrium values 
(possibly due to the wall layer) can cause larger errors in the density and viscosity calculations
because the density and viscosity are dependent on $\phi$
through some linear functions or functions of some other suitable forms
(e.g., see~\cite{pre07-geom-wbc, jcp10lbm-drop-impact}).
Base on such considerations, it seems that CubSE and Geom
are indeed more preferable than LinSE for PF simulations involving contact lines on a
lyophobic or lyophilic wall (i.e., $\theta_{w} \neq 90^{\circ}$).

\begin{figure}[htp]
  \centering
  	(a) \includegraphics[scale = 0.60]{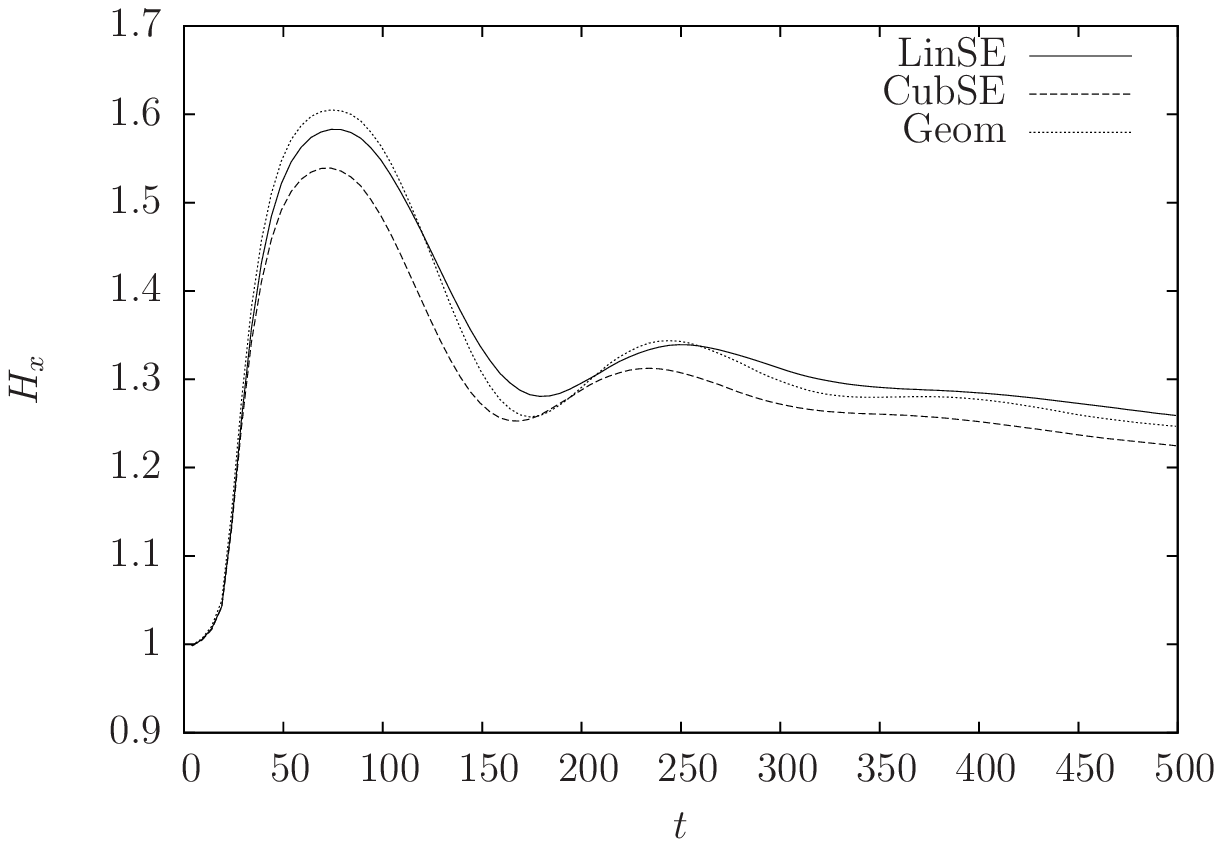}	
	(b) \includegraphics[scale = 0.60]{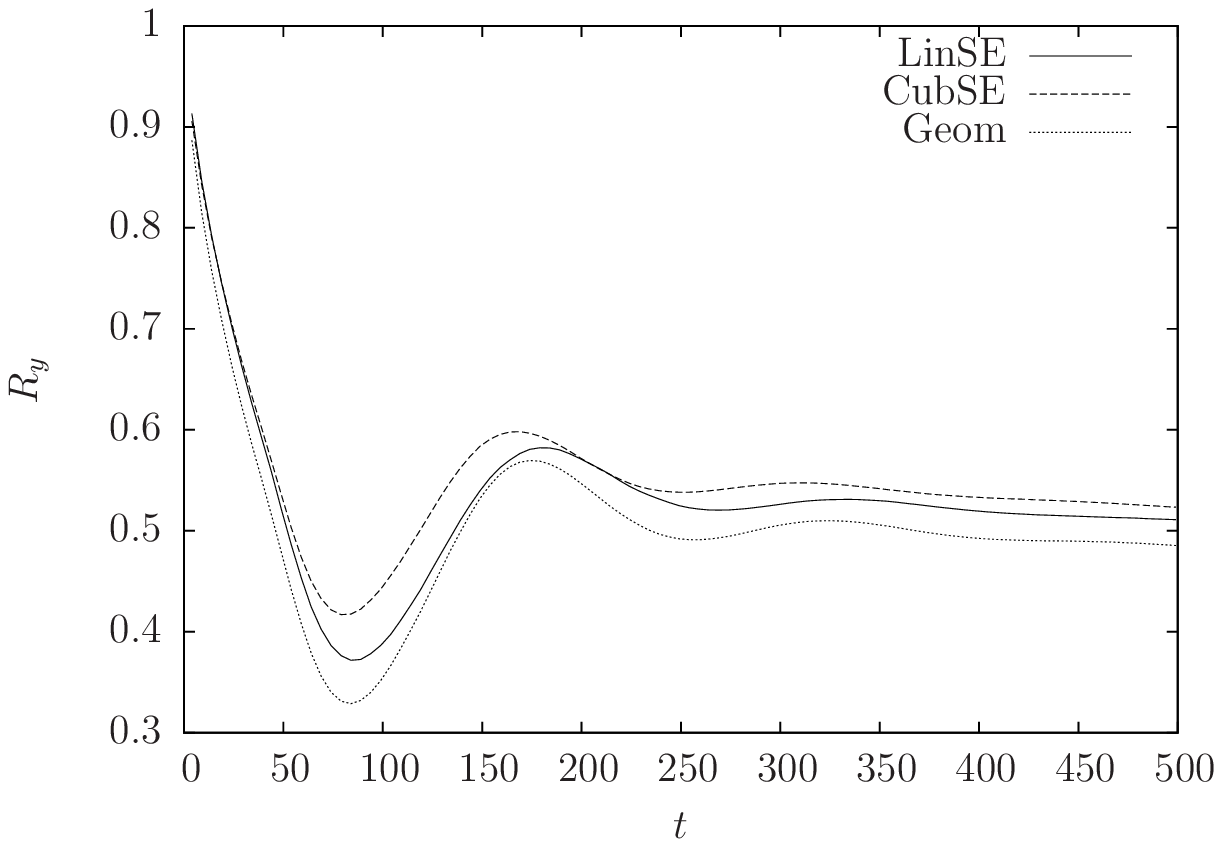}
        (c) \includegraphics[scale = 0.58]{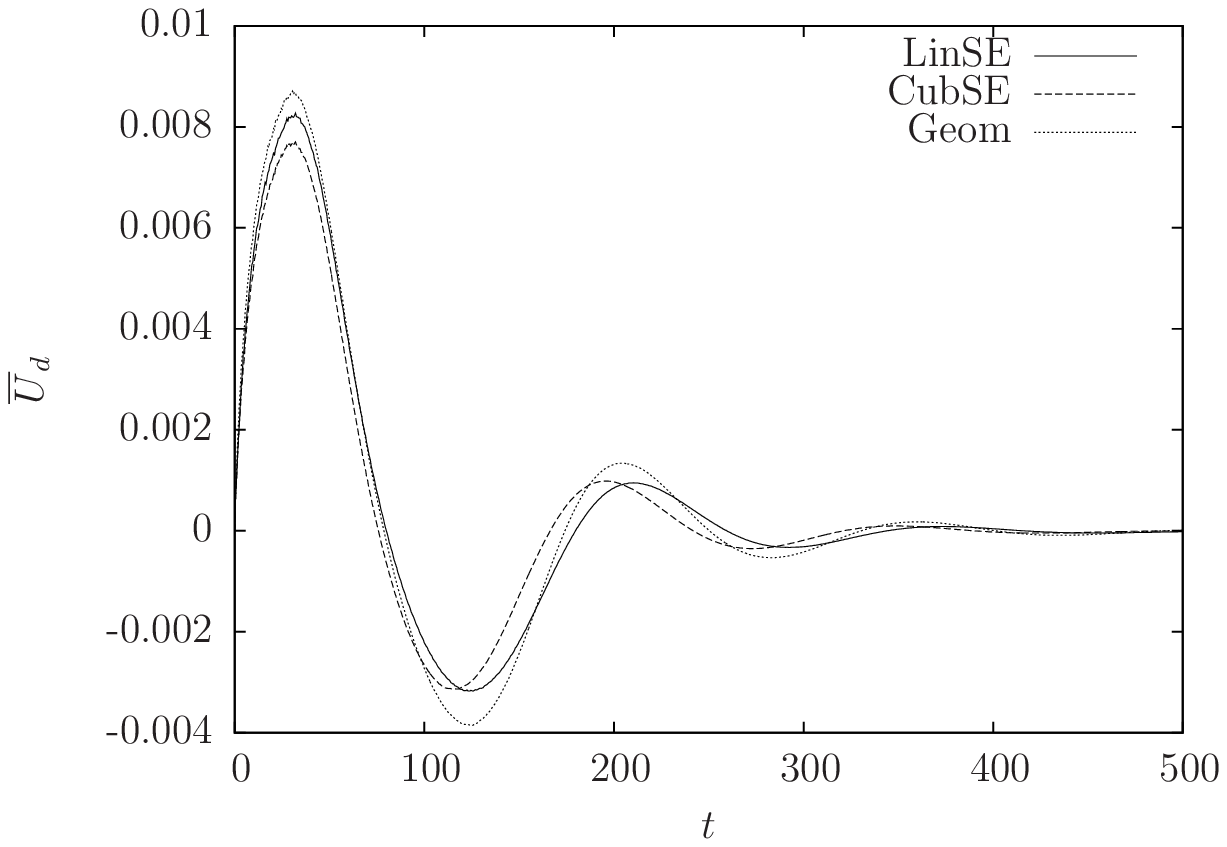}
        (d) \includegraphics[scale = 0.58]{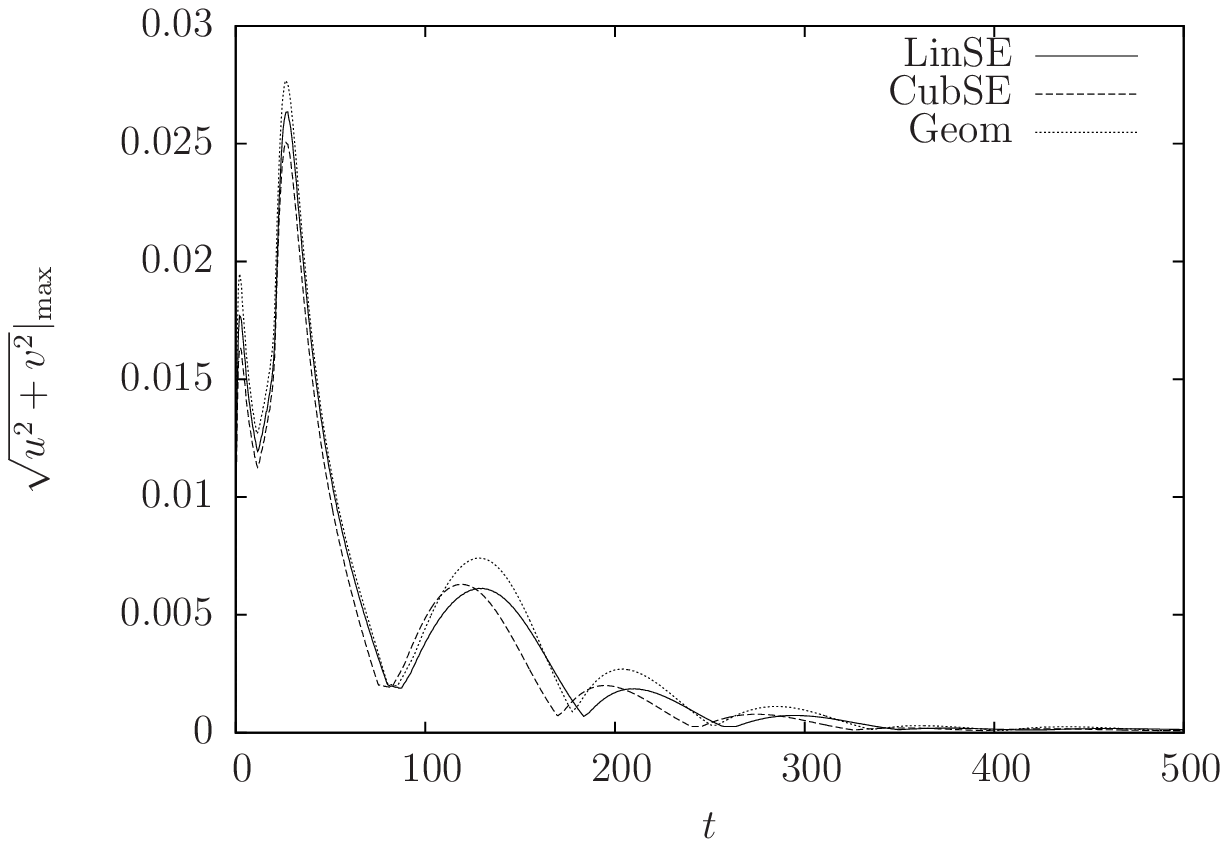}
  \caption{Evolutions of 
    (a) the drop \emph{height} on the $x-$axis $H_{x}$
 (b) the drop \textit{radius} on the left wall $R_{y}$  
  (c) the average drop velocity $\overline{U}_{d}$ in the $x-$direction 
  (d) the maximum velocity magnitude $\sqrt{u^2+v^2}\vert _{\textrm{max}}$
   by using different WBCs.
  The parameters are $\theta_{w} = 135^{\circ}$, $\theta_{i} = 90^{\circ}$, 
  $Re=1000$ ($Oh = 0.032$), $Ca=1$,
  $Cn=0.1$, $Pe=5000$ ($S = 0.014$), $N_{L} = 40$, $N_{t} = 240$.}
  \label{fig:cmp-ud-vm-Re1k-ca135-cn0d1-Pe5k}
\end{figure}

Next we examine the drop \emph{height} $H_{x}$ along the $x-$axis,
the drop \textit{radius} on the left wall $R_{y}$,
the average drop velocity $\overline{U}_{d}$ in the $x-$direction,
and the maximum velocity magnitude $\sqrt{u^2+v^2}\vert_{\textrm{max}}$
for this case.
Figure \ref{fig:cmp-ud-vm-Re1k-ca135-cn0d1-Pe5k} compares
the evolutions of these four quantities obtained by using LinSE, CubSE and Geom.
It is observed from Fig. \ref{fig:cmp-ud-vm-Re1k-ca135-cn0d1-Pe5k} 
that all WBCs predict similar trends for all these quantities,
and the evolutions of the $\overline{U}_{d}$ and $\sqrt{u^2+v^2}\vert _{\textrm{max}}$
by different WBCs are closer to each other than those of $H_{x}$ and $R_{y}$.
This could be due to that $H_{x}$ and $R_{y}$ are two local quantities 
and more sensitive to the change of WBC.
Among all WBCs, Geom gives the strongest oscillations (with the largest amplitudes)
for all the four quantities, whereas CubSE predicts the weakest oscillations.
The results by LinSE are located somewhere in between.
This is similar to the observation reported in Section \ref{ssec:lc-wg}
for a liquid column driven by WG.
At the same time, the periods of oscillation for $H_{x}$ and $R_{y}$
predicted by LinSE are slightly larger.
Since by using Geom one always imposes $\theta_{w}$ exactly 
(for instance, at the beginning the interface near the contact line is suddenly \textit{bent}
from $\theta_{i}=90^{\circ}$ to match the given one $\theta_{w}=135^{\circ}$),
it is not hard to understand that under this condition
the drop dewets more violently and shows larger oscillations.
It seems that less surface energy
was converted into kinetic energy when CubSE was applied.
The reason remains to be explored.

\begin{figure}[htp]
  \centering
  \includegraphics[scale = 0.8]{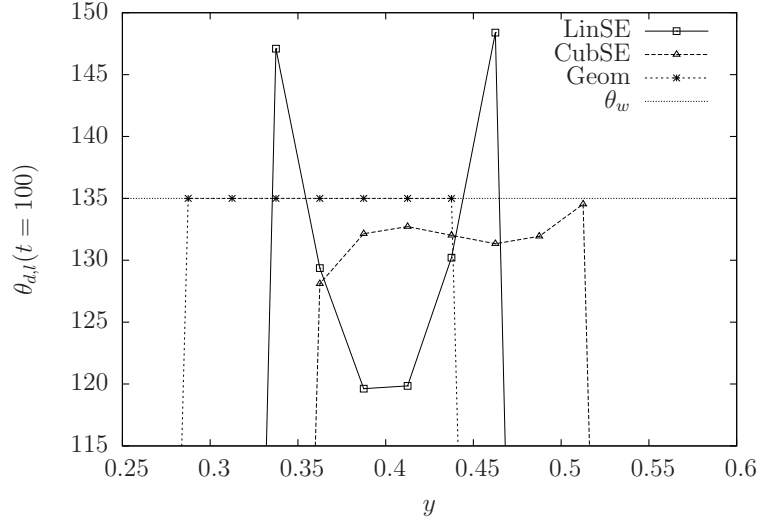}
  \caption{Comparison of the local contact angle $\theta_{d, l}$ 
  in the interfacial region near the contact line at $t=100$ by using different WBCs.
  The parameters are $\theta_{w} = 135^{\circ}$, $\theta_{i} = 90^{\circ}$, 
  $Re=1000$ ($Oh = 0.032$), $Ca=1$,
  $Cn=0.1$, $Pe=5000$ ($S = 0.014$), $N_{L} = 40$, $N_{t} = 240$.
The horizontal dashed line corresponds to $\theta_{w}$.}
  \label{fig:cmp-lca-Re1k-ca135-cn0d1-Pe5k}
\end{figure}

\begin{figure}[htp]
  \centering
	(a) \includegraphics[scale = 0.22]{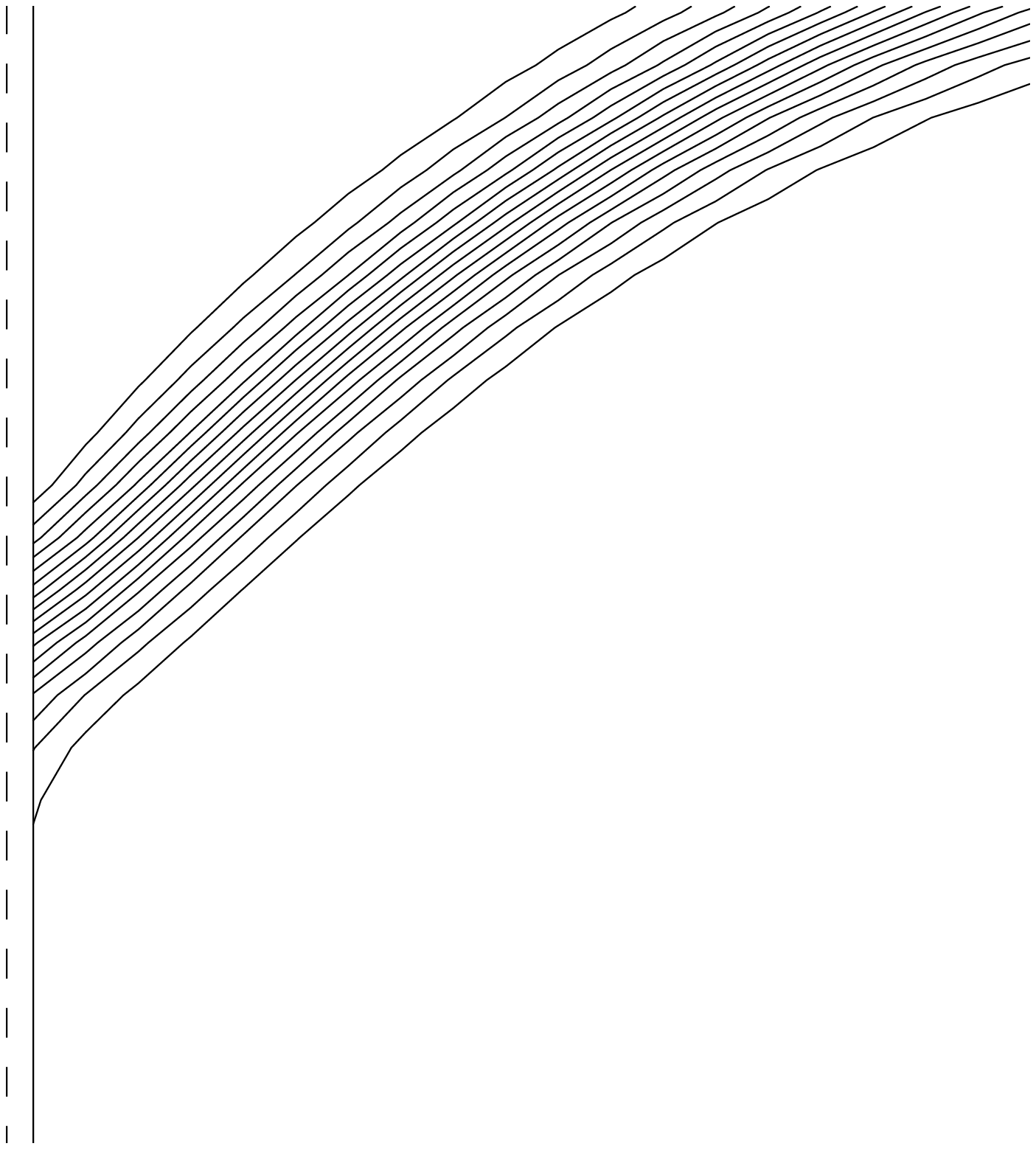}
	(b) \includegraphics[scale = 0.22]{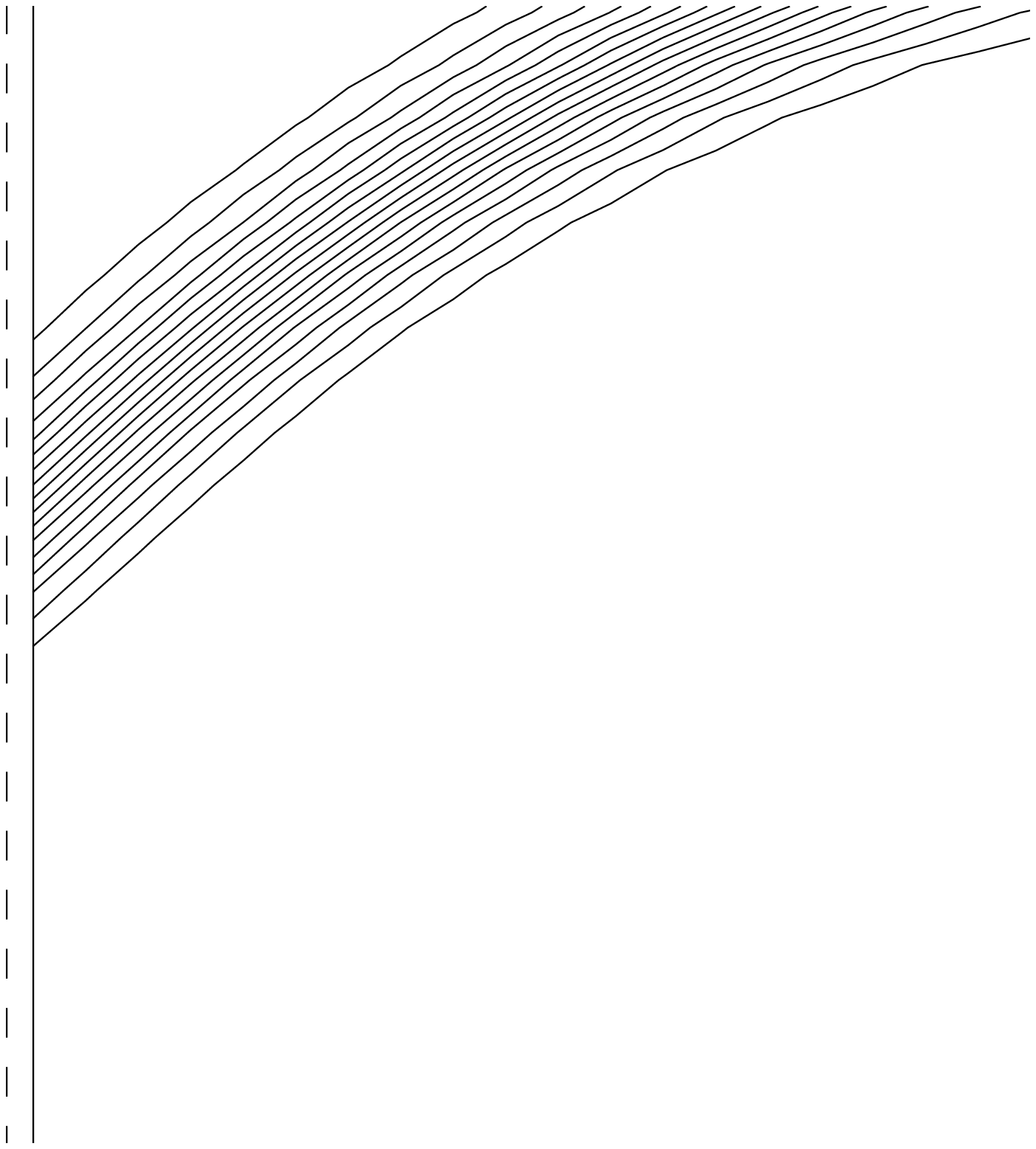}
	(c) \includegraphics[scale = 0.22]{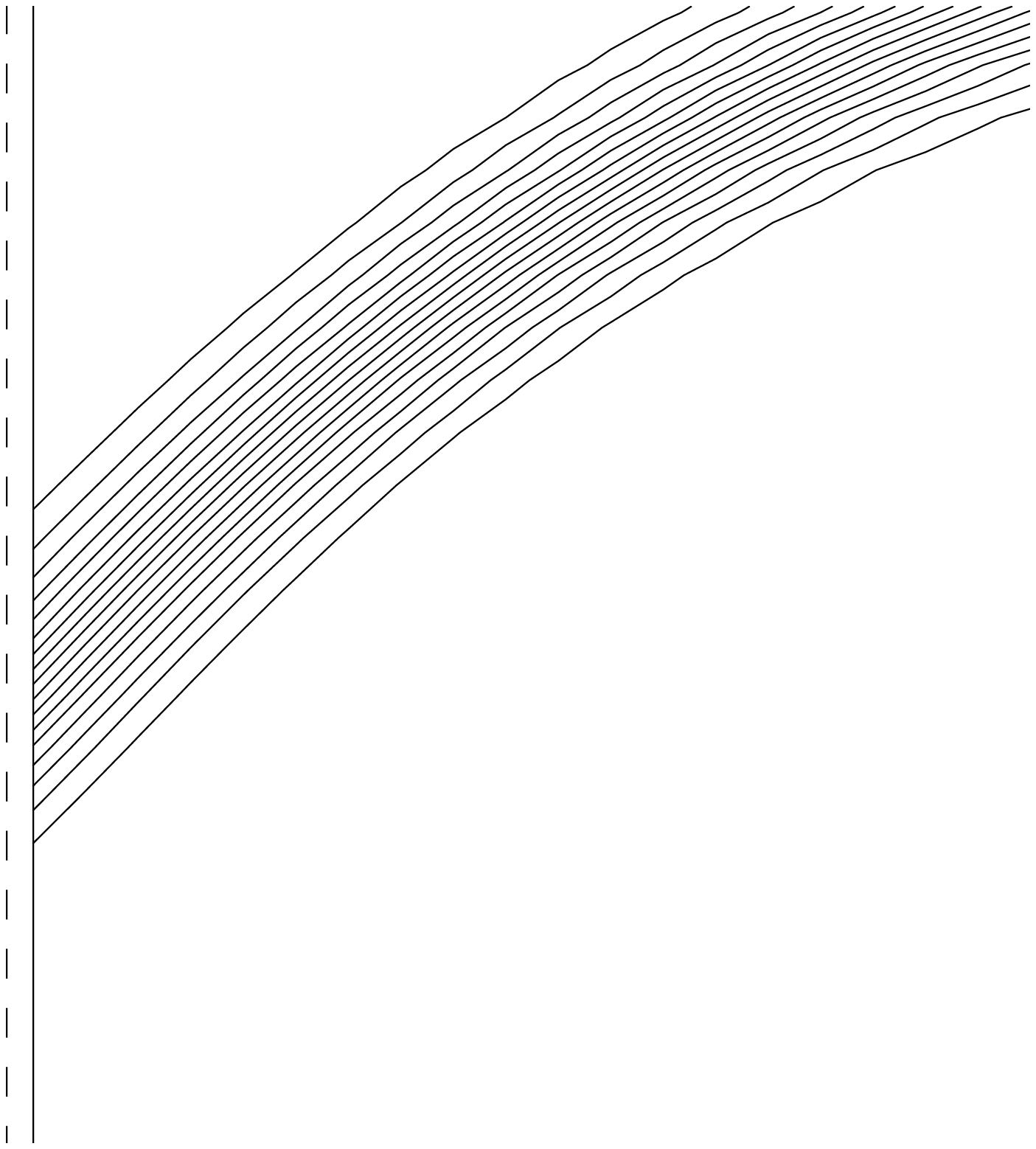}	
  \caption{Comparison of the contours of $\phi$ 
  	in the interfacial region near the contact line at $t=100$ by using different WBCs:
	(a) LinSE; (b) CubSE; (c) Geom.
  The parameters are $\theta_{w} = 135^{\circ}$, $\theta_{i} = 90^{\circ}$, 
  $Re=1000$ ($Oh = 0.032$), $Ca=1$,
  $Cn=0.1$, $Pe=5000$ ($S = 0.014$), $N_{L} = 40$, $N_{t} = 240$.
  The left dashed line represents the \textit{actual} location of the wall,
  which is $0.5 h$ away from the left boundary shown here.}
  \label{fig:2d-cmp-lca-Re1k-ca135-cn0d1-Pe5k}
\end{figure}

\begin{figure}[htp]
  \centering
  \includegraphics[scale = 0.8]{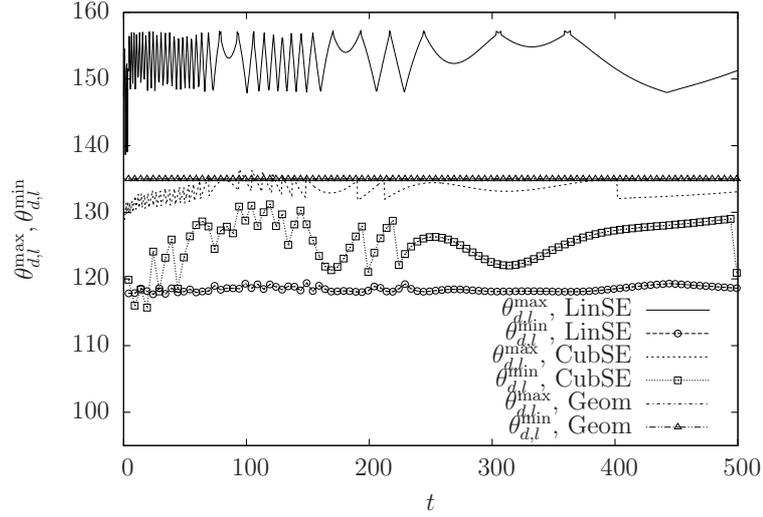}
  \caption{Comparison of the evolutions of the maximum
  and minimum local contact angles across the interfacial region
  near the contact line,
   $\theta_{d, l}^{\textrm{max}}$  and $\theta_{d, l}^{\textrm{min}}$, 
  by using different WBCs.
  The parameters are $\theta_{w} = 135^{\circ}$, $\theta_{i} = 90^{\circ}$, 
  $Re=1000$ ($Oh = 0.032$), $Ca=1$,
  $Cn=0.1$, $Pe=5000$, $N_{L} = 40$, $N_{t} = 240$.}
  \label{fig:cmp-evol-lca-Re1k-ca135-cn0d1-Pe5k}
\end{figure}

In Section \ref{sssec:wbc-geom}, when deriving Geom,
it was assumed that the contours of the order parameter $\phi$ are parallel to each other
across the interface near the contact line.
In Section \ref{ssec:char-quant-diml-num-setup},
a local contact angle $\theta_{d, l}$ was calculated from the local gradients
of $\phi$ in the interfacial region (c.f. Eq. (\ref{eq:LocalCA})).
Now we examine $\theta_{d, l}$ when different WBCs are used for the above case.
Figure \ref{fig:cmp-lca-Re1k-ca135-cn0d1-Pe5k} 
compares $\theta_{d, l}$ in the interfacial region at $t=100$ for the three WBCs.
Note that the tangential gradient in Eq. (\ref{eq:LocalCA})
was exactly on the wall (obtained from extrapolations like Eq. (\ref{eq:tang-grad-extrap}))
and the normal gradient was also exactly on the wall
(calculated by finite difference scheme using $\phi$ 
located in the two neighboring layers that are $0.5 h$ away from the wall).
It is found from Fig. \ref{fig:cmp-lca-Re1k-ca135-cn0d1-Pe5k} 
that, except when Geom is applied,
$\theta_{d, l}$ in the interfacial region
shows some deviations from $\theta_{w}$.
The deviations are relatively small (mostly less than $5^{\circ}$) when CubSE is applied.
But when LinSE is used, large fluctuations (with deviations as large as $15^{\circ}$) 
are seen in $\theta_{d, l}$.
This indicates that the contours of $\phi$ are no longer parallel to each other,
which is confirmed by Fig. \ref{fig:2d-cmp-lca-Re1k-ca135-cn0d1-Pe5k},
which compares such contours near the contact line 
at the selected moment ($t=100$) for the three WBCs.
At the same time, Fig. \ref{fig:2d-cmp-lca-Re1k-ca135-cn0d1-Pe5k}
shows that when CubSE or Geom is applied,
the contours of $\phi$ appear to be indeed parallel.
The large fluctuations of $\theta_{d, l}$ by using LinSE are very likely to be related
to the wall layer, which may distort the profile of $\phi$ near the contact line.
Besides the particular moment $t=100$, the maximum and minimum
values of $\theta_{d, l}$ in the interfacial region during the whole simulation
are shown in Fig. \ref{fig:cmp-evol-lca-Re1k-ca135-cn0d1-Pe5k}.
It is seen that $\theta_{d, l}$ is always equal to $\theta_{w}$ when Geom is used.
When LinSE is applied, $\theta_{d, l}^{\textrm{max}}$ itself shows large fluctuations
during the simulation (especially in the early stage)
whereas $\theta_{d, l}^{\textrm{min}}$ fluctuates much less violently.
When CubSE is applied, $\theta_{d, l}$ is almost always smaller than $\theta_{w}$,
and for most of the time,
$\theta_{d, l}^{\textrm{max}}$ and $\theta_{d, l}^{\textrm{min}}$ 
are both within the band delimited by $\theta_{w}$ and the minimum value of $\theta_{d, l}$
when LinSE is used.

\begin{figure}[htp]
  \centering
(a)  \includegraphics[scale = 0.58]{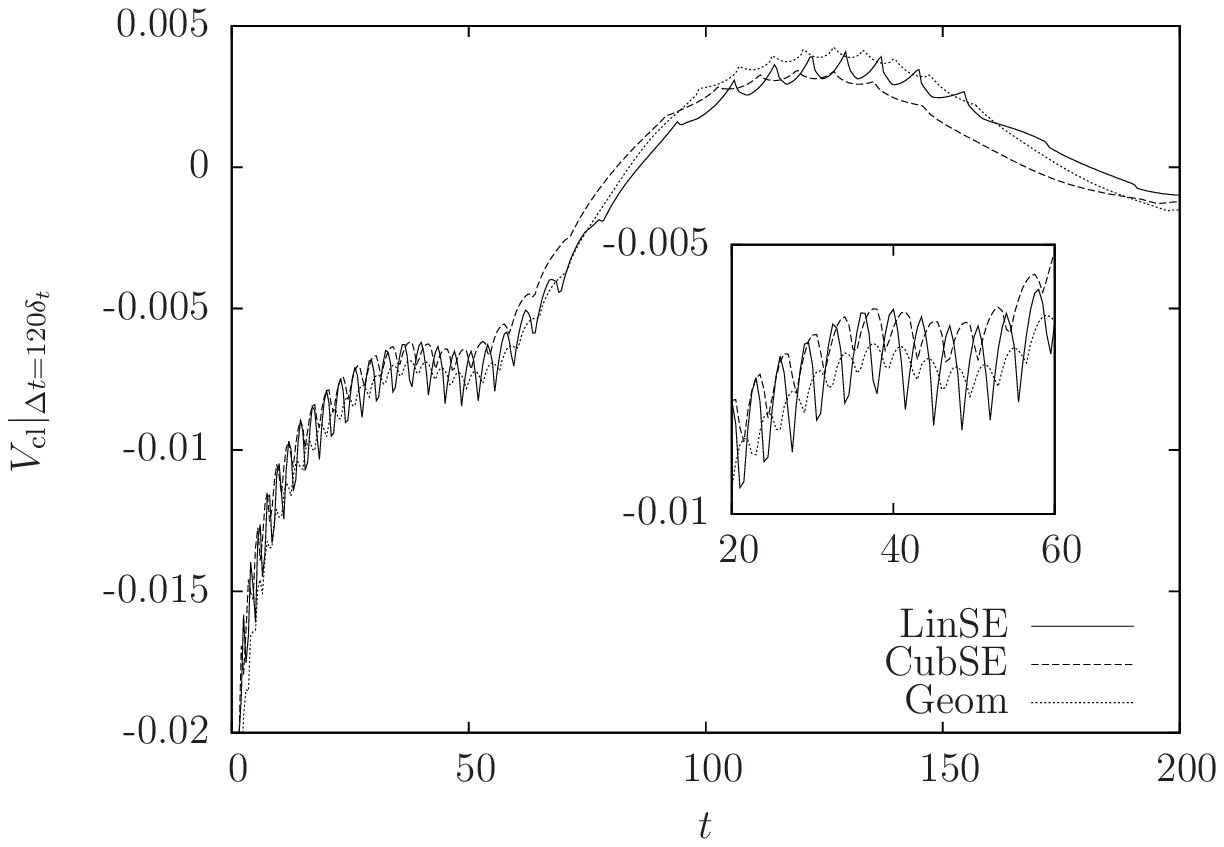}
(b)  \includegraphics[scale = 0.58]{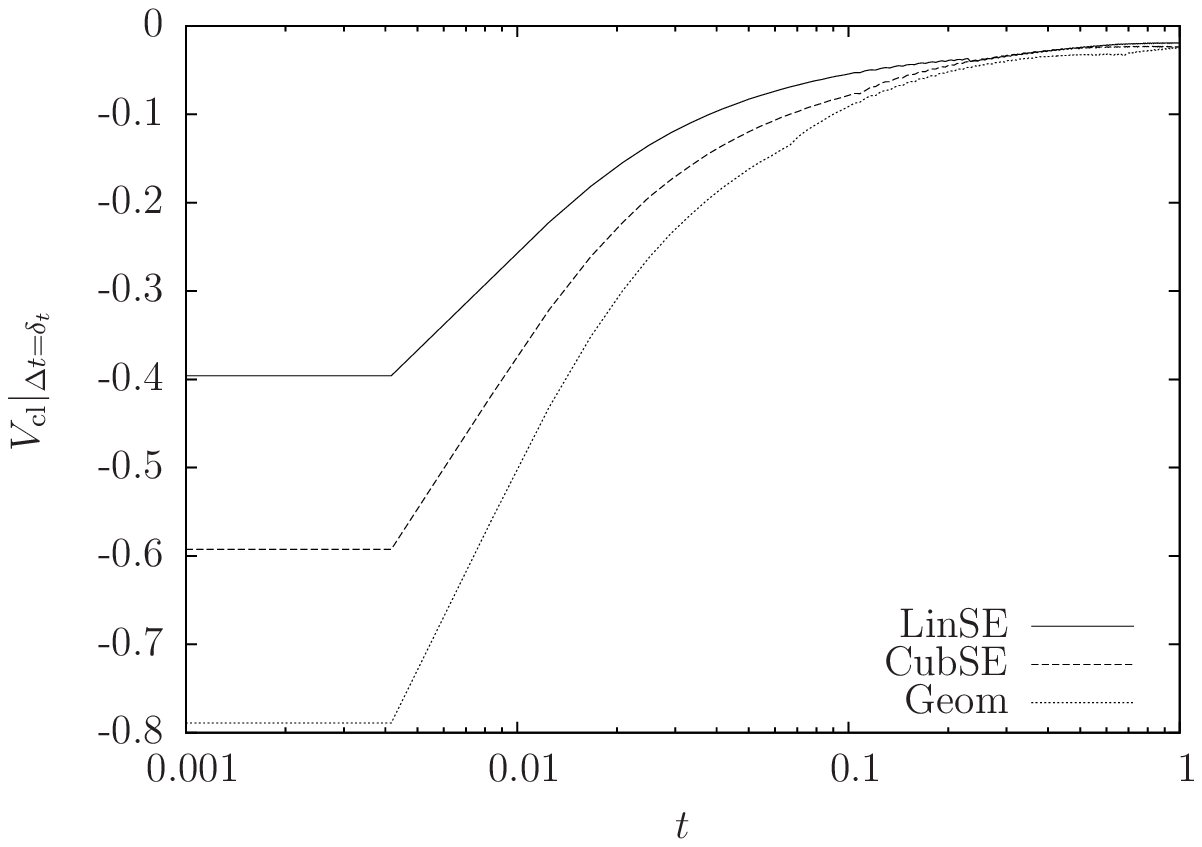}
  \caption{Comparison of the evolutions of the contact line velocity $V_\textrm{cl}$ 
  (a) in $0\leq t < 200$ obtained at $\Delta t =  120 \delta_{t}$ (equivalent to $0.5 T_{c}$)
  (b) in $0\leq t < 1$ (with $t$ plotted in log scale) obtained at $\Delta t = \delta_{t}$ 
  by using different WBCs.
  The parameters are $\theta_{w} = 135^{\circ}$, $\theta_{i} = 90^{\circ}$, 
  $Re=1000$ ($Oh = 0.032$), $Ca=1$,
  $Cn=0.1$, $Pe=5000$ ($S = 0.014$), $N_{L} = 40$, $N_{t} = 240$.}
  \label{fig:cmp-vcl-te100-te1-Re1k-ca135-cn0d1-Pe5k}
\end{figure}

In addition to the above quantities, we also studied 
the contact line velocity $V_\textrm{cl}$ when different WBCs were applied.
Figure \ref{fig:cmp-vcl-te100-te1-Re1k-ca135-cn0d1-Pe5k} 
compares
the evolutions of two contact line velocities:
one obtained at each half characteristic time $T_{c}$ 
($\Delta t =  120 \delta_{t}$, see Fig. \ref{fig:cmp-vcl-te100-te1-Re1k-ca135-cn0d1-Pe5k}a)
and the other obtained at each time step 
($\Delta t = \delta_{t}$, see Fig. \ref{fig:cmp-vcl-te100-te1-Re1k-ca135-cn0d1-Pe5k}b).
Based on Fig. \ref{fig:cmp-vcl-te100-te1-Re1k-ca135-cn0d1-Pe5k}a,
it is observed that over a relatively large time scale ($0\leq t < 200$)
the contact line velocities $V_\textrm{cl}\vert_{\Delta t =  120 \delta_{t}}$
by using different WBCs are close to each other,
and oscillations are observed in $V_\textrm{cl}\vert_{\Delta t =  120 \delta_{t}}$
for all the WBCs 
in the early stage ($0\leq t < 75$, see the inset in 
Fig. \ref{fig:cmp-vcl-te100-te1-Re1k-ca135-cn0d1-Pe5k}a
for a closer view of $20\leq t \leq 60$).
At the same time, it is seen that when LinSE is used,
$V_\textrm{cl}\vert_{\Delta t =  120 \delta_{t}}$
oscillates at a relatively large amplitude
as compared to those obtained by using the other two WBCs.
This could be attributed to the wall layer and the high non-uniformity of the local contact angle
across the interfacial region near the contact line discussed above.
On the other hand, at the very early stage ($0\leq t < 1$)
over a much smaller time scale, 
the contact line velocities $V_\textrm{cl}\vert_{\Delta t =  \delta_{t}}$ 
appear to be \textit{much smoother}
under all WBCs, as found from 
Fig. \ref{fig:cmp-vcl-te100-te1-Re1k-ca135-cn0d1-Pe5k}b.
Besides, 
one finds from Fig. \ref{fig:cmp-vcl-te100-te1-Re1k-ca135-cn0d1-Pe5k}b 
that $V_\textrm{cl}\vert_{\Delta t =  \delta_{t}}$ at the beginning ($t=0$)
obtained with Geom has the largest magnitude (about $0.8$)
whereas that obtained with LinSE  has the smallest magnitude (about $0.4$),
and that with CubSE lies in between (about $0.6$).
This is somewhat different from previous observations of the evolutions
of $R_{y}$ and $\overline{U}_{d}$ over a larger time scale.
In that situation, as discussed above, the motions of the contact line
and the drop were the weakest when CubSE was applied.
Such differences between the very early stage and the later stage for different WBCs
are not yet known and requires further study.

\section{Concluding Remarks}\label{sec:conclusion}

To summarize, we have examined five different wetting boundary conditions, 
using the linear, cubic and sine forms of surface energy (LinSE, CubSE and SinSE), 
the geometric formulation (Geom) and the characteristic interpolation (CI) respectively,
in phase-field-based simulations of several drop problems.
It has been found that they may be categorized into three groups:
(1) LinSE; (2) CubSE and SinSE; (3) Geom and CI.
Among each of the latter two groups, 
the WBCs gave similar (or even nearly identical) predictions 
of key observables of the flow.
While it may lead to different results for capillarity-driven flows by using different WBCs,
all WBCs had very close predictions for the mechanically-driven liquid column.
For the WG-driven liquid column, Geom did the best in terms of
the consistency in the predicted drop velocity and dynamic contact angles,
whereas LinSE was the worst in this aspect.
For drop dewetting, several local and average quantities were examined
under different WBCs.
The drop underwent stronger motions during the dewetting process with Geom than with CubSE applied, 
and the initial contact line velocity was found to be dependent on many factors, 
including 
the WBC, as well as the Cahn number and Peclet number.
When LinSE was employed for a wall with a contact angle not equal to $90^{\circ}$, 
a wall layer appeared and the local contact angle near the contact line showed large fluctuations. 
Although LinSE has deficiencies mentioned above,
it gave results on many quantities of interest
somewhere in between those by the other two groups.
When its extreme simplicity is further considered,
LinSE might still be regarded as a good candidate for preliminary studies.
In addition to the comparisons, 
a simple procedure to mimic the wall energy relaxation
has been proposed based on the hybridization of the existing WBCs 
with another implementation of complete hysteresis.
This simple additional step can significantly improve the agreement
between the simulated results and the experimental data
when the hybridization parameter is tuned properly.
This study not only discloses the dynamics and certain characteristics of 
several basic drop problems, 
but may also provide useful guidelines on the choice of the WBC
and some new valuable development for future work. 

\bf Acknowledgement \rm

This work is supported by Natural Science Foundation Project
of CQ CSTC No. 2011BB6078,
the National Natural Science Foundation of China 
(NSFC, Grant No. 11202250) 
and the Fundamental Research Funds for the Central Universities
(Project No. CDJZR12110001, CDJZR12110072).

\end{document}